\documentclass[aps,prb,reprint,superscriptaddress,longbibliography,floatfix]{revtex4-2}

\usepackage{upgreek}
\usepackage[version=3]{mhchem} 
\usepackage{amsmath,amssymb}
\usepackage{cancel}
\usepackage{bm}
\usepackage{subfigure}
\usepackage{xcolor}
\colorlet{BLUE}{blue}
\renewcommand{\selectlanguage}[1]{}
\usepackage{comment}
\usepackage[colorlinks=true, citecolor=blue, linkcolor=blue,urlcolor=blue]{hyperref}
\usepackage{todonotes}
\usepackage{tikz}
\usepackage{enumitem}
\usepackage{tabularx}
\usepackage{array}
\usepackage{ragged2e}
\usepackage{booktabs}
\usetikzlibrary{patterns,positioning,arrows.meta,decorations.pathreplacing}
\newcommand*{\tauB}{\tau_\mathrm{B}}  
\newcommand*{\tauN}{\tau_\mathrm{N}}  
\newcommand*{\tauD}{\tau_\mathrm{D}}  
\newcommand*{\br}{\vec{r}}      
\newcommand*{\dH}{d_\mathrm{H}}   
\newcommand*{\baxis}{\vec{e}}      
\newcommand*{\bmag}{\vec{m}}      
\newcommand*{\bF}{\vec{F}}      
\newcommand*{\bT}{\vec{T}}      
\newcommand*{\bv}{\vec{v}}      
\newcommand*{\bUpsilon}{\boldsymbol{\Upsilon}} 
\newcommand*{\bomega}{\vec{\omega}} 
\newcommand*{\bnabla}{\vec{\nabla}} 

\newcommand*{\eav}{\langle \vec{{m}} \rangle} 
\newcommand*{\bxi}{\vec{\xi}} 

\newcolumntype{Y}{>{\RaggedRight\arraybackslash}X}
\newcolumntype{C}{>{\Centering\arraybackslash}X}

\definecolor{micromagnetic_purple}{RGB}{128, 0, 128}
\definecolor{angular_velocity_teal}{RGB}{0, 128, 128}

\begin{document}

\title{Perspective on Magnetic Nanoparticle Modeling: Interactions, Timescales and Regimes}
\author{Deniz Mostarac}
\affiliation{%
School of Chemistry, University of Edinburgh, EH9 3FJ Edinburgh, United Kingdom}%
\email{deniz.mostarac@ed.ac.uk}
\author{Andrey A. Kuznetsov}
\affiliation{%
Faculty of Physics, University of Vienna, 1090 Vienna, Austria}
\author{Manuel Wolfschwenger}
\affiliation{%
Institute of Electrical and Biomedical Engineering, UMIT TIROL– Private University
for Health Sciences and Health Technology, 6060 Hall in Tirol, Austria}%
\author{Santiago Helbig}
\affiliation{%
Faculty of Physics, University of Vienna, 1090 Vienna, Austria}%
\author{Claas Abert}
\affiliation{%
Faculty of Physics, University of Vienna, 1090 Vienna, Austria}%
\author{Rudolf Weeber}
\affiliation{%
Institute for Computational Physics, University of Stuttgart, Stuttgart, Germany}%
\author{Daniel Baumgarten}
\affiliation{%
Biomedical Engineering Group, Department of Mechatronics, University of Innsbruck, 6020 Innsbruck, Austria}%
\author{Patrick Ilg}
\affiliation{%
School of Mathematical, Physical and Computational Sciences, University of Reading, RG6 6ED Reading, United Kingdom}%
\author{Dieter Suess}
\affiliation{%
Faculty of Physics, University of Vienna, 1090 Vienna, Austria}%
\author{Sofia Kantorovich}
\affiliation{%
Faculty of Physics, University of Vienna, 1090 Vienna, Austria}%



\begin{abstract}
The response of magnetic nanoparticles (MNPs) to applied magnetic fields underpins a broad range of biomedical and technological applications. In this Perspective, we review the principal modeling approaches for describing MNP dynamics across different physical regimes, ranging from coarse-grained macrospin descriptions to spatially resolved micromagnetic simulations. Selecting an appropriate model depends on the relevant energy scales and timescales, including those associated with magnetic anisotropy. We compare the assumptions, computational requirements, and regimes of applicability of the fixed-point-dipole, effective-field, thermal Stoner--Wohlfarth, diffusion-jump, coupled Landau--Lifshitz--Gilbert, egg, and micromagnetic models. Particular attention is given to coupling magnetization dynamics with translational and rotational particle motion, hydrodynamic interactions, and long-range dipolar interactions. By relating the relevant physical regimes to the resolution and computational cost of each approach, we provide practical guidance for model selection and outline challenges for predictive multiscale simulations of interacting MNP systems.
\end{abstract}

\maketitle

\setcounter{tocdepth}{3}
\tableofcontents

\vspace{20pt}

\section{Introduction}\label{sec::intro}
Magnetic nanoparticles (MNPs) have attracted considerable interest because their distinctive physical properties and controllable response to externally applied magnetic fields enable a wide range of applications in both biomedical and technical fields~\cite{materon2021magnetic, socoliuc_ferrofluids_2022}. In biomedical contexts, MNPs are used for targeted drug delivery~\cite{Pri18}, gene therapy~\cite{Maj16}, magnetic hyperthermia~\cite{Ort13}, and for imaging modalities such as magnetorelaxometry imaging (MRXI)~\cite{Jau24, Ars23} and magnetic particle imaging (MPI)~\cite{Coe22}. Their controlled response to externally applied magnetic fields allows for precise localization and manipulation, which is essential for minimally invasive diagnostics and therapeutic procedures~\cite{materon2021magnetic}. Systems employed for biomedical applications are sometimes even referred to as bio-ferrofluids~\cite{socoliuc_ferrofluids_2022}. Beyond biomedical use, MNPs are also increasingly employed in technical applications, including sensing~\cite{Glo19}, magnetic bearings~\cite{Pat22}, wastewater treatment~\cite{Ali17} and magnetic separation processes~\cite{Sim18,wolfschwenger2025dual}. The reliable design and operation of such applications require a detailed understanding of a variety of complex and coupled phenomena. These include particle–particle interactions, hydrodynamic interactions, interactions between MNPs and surrounding structures or biological media, changes in ferrofluid viscosity under applied magnetic fields, commonly referred to as the magnetoviscous effect~\cite{Ode02}, the formation of particle aggregates and internal single particle magnetization dynamics. In addition, the performance of a given application is often highly sensitive to flow-related characteristics, including diffusion coefficients and viscosity. For example, in magnetic drug targeting, accurate knowledge of these quantities is essential to predict the number of particles that can be accumulated locally, which constitutes the long-term objective~\cite{Lin21}. To support the development and optimization of both biomedical and technical applications, computational and mathematical models are essential for understanding these coupled mechanisms. Furthermore, they provide predictive capabilities for estimating system behavior. They also guide the selection and optimization of application-specific parameters~\cite{Ilg_lnp}.

For this purpose, analytical and numerical macroscopic models have been formulated to characterize how ferrofluids behave when subjected to externally applied magnetic fields. Two fundamental approaches are commonly used to model ferrofluid behavior: the Lagrangian approach~\cite{For18, Sha19, Lin21}, which computes the trajectories of individual particles, and the Eulerian approach~\cite{Gon21, Leo15, Ken16}, which describes the evolution of particle concentration within a control volume. By tracking individual particles, Lagrangian models can capture microscopic phenomena, including interparticle interactions and the internal magnetization dynamics of individual particles. However, explicitly resolving the relevant microscopic processes, which span a broad range of length scales and timescales, can incur a tremendous computational cost. Eulerian models address this limitation by treating the suspension as a continuum rather than resolving individual particles. As a result, microscopic effects are not represented directly and must instead be incorporated through constitutive equations. The predictive accuracy of such macroscopic models therefore depends strongly on the availability and reliability of these constitutive relations, which are often not known a priori~\cite{ilg_structure_2006}.
Therefore, the underlying effects must first be investigated at the microscopic, single-particle level. This has motivated the development of a broad range of microscopic and micromagnetic models tailored to specific particle properties, interaction mechanisms, and computational constraints.\\
In this perspective, we review these models, discuss their regimes of applicability, and assess their respective advantages and limitations. We also provide an outlook on open challenges and propose directions that we consider particularly important for the future development of the field.

\section{Key Energy and Time Scales}\label{sec::relevant_scales}

\begin{table*}[tp]
\scriptsize
\centering
\caption{Summary of the symbols used throughout this work, with their definition and SI unit. A dash ($-$) denotes a dimensionless quantity. An arrow ($\vec{\ }$) denotes a vector and bold ($\bm{N}$) a tensor or matrix.}
\label{tab:nomenclature}
\setlength{\tabcolsep}{4pt}
\renewcommand{\arraystretch}{1.1}
\begin{tabular}{@{}lll@{\hskip 1em}@{\color{black!20}\vrule}@{\hskip 1em}lll@{}}
\hline
Symbol & Definition & Unit & Symbol & Definition & Unit \\
\hline
\multicolumn{6}{@{}l}{\textbf{Magnetization, moment, and orientation}}\\
\noalign{\vskip 2pt}
$\bmag$ & normalized magnetization & $-$ & $\baxis$ & easy-axis direction & $-$ \\
$\vec{M}$ & magnetization & A\,m$^{-1}$ & $M_s$ & saturation magnetization & A\,m$^{-1}$ \\
$\vec{\mu}$ & magnetic moment & A\,m$^{2}$ & $\mu$ & magnetic moment magnitude & A\,m$^{2}$ \\
\noalign{\vskip 6pt}
\multicolumn{6}{@{}l}{\textbf{Magnetic fields and tensors}}\\
\noalign{\vskip 2pt}
$\vec{H}_\text{ext}$ & applied (external) field & A\,m$^{-1}$ & $\vec{H}_\text{eff}$ & effective field & A\,m$^{-1}$ \\
$\vec{H}_d$ & demagnetization (stray) field & A\,m$^{-1}$ & $\vec{H}_\text{ani}$ & anisotropy field & A\,m$^{-1}$ \\
$\vec{H}_\text{therm}$ & thermal (stochastic) field & A\,m$^{-1}$ & $\vec{H}_{\text{Bar}}$ & Barnett field & A\,m$^{-1}$ \\
$\vec{H}_E$ & energy-derived field & A\,m$^{-1}$ & & & \\
$\widetilde{\bm{N}}$ & demagnetization tensor & m$^{-3}$ & & & \\
\noalign{\vskip 6pt}
\multicolumn{6}{@{}l}{\textbf{Magnetization dynamics}}\\
\noalign{\vskip 2pt}
$\alpha$ & Gilbert damping parameter & $-$ & $\gamma$ & gyromagnetic ratio & s$^{-1}$\,T$^{-1}$ \\
$\zeta^m$ & magnetic friction coefficient & J\,s & $\eta^m$ & magnetic viscosity & Pa s \\
\noalign{\vskip 6pt}
\multicolumn{6}{@{}l}{\textbf{Material and physical constants}}\\
\noalign{\vskip 2pt}
$A$ & exchange stiffness constant & J\,m$^{-1}$ & $K$ & uniaxial anisotropy constant & J\,m$^{-3}$ \\
$\mu_0$ & vacuum permeability & T\,m\,A$^{-1}$ & $\rho$ & particle mass density & kg\,m$^{-3}$ \\
$\eta$ & carrier-fluid viscosity & Pa\,s & $k_B$ & Boltzmann constant & J\,K$^{-1}$ \\
$T$ & temperature & K & & & \\
\noalign{\vskip 6pt}
\multicolumn{6}{@{}l}{\textbf{Energies and pair potentials}}\\
\noalign{\vskip 2pt}
$E$ & total magnetic energy & J & $U$ & potential energy & J \\
$u^\mathrm{LJ}$ & Lennard-Jones potential & J & $u^\mathrm{WCA}_{ij}$ & Weeks--Chandler--Andersen potential & J \\
$\varepsilon$ & Lennard-Jones well depth & J & $u_{ij}^\mathrm{dd}$ & dipole--dipole pair energy & J \\
\noalign{\vskip 6pt}
\multicolumn{6}{@{}l}{\textbf{Dimensionless parameters}}\\
\noalign{\vskip 2pt}
$\xi$ & Langevin parameter & $-$ & $\sigma$ & anisotropy parameter & $-$ \\
$\lambda$ & dipolar coupling parameter & $-$ & $L$ & Langevin function & $-$ \\
\noalign{\vskip 6pt}
\multicolumn{6}{@{}l}{\textbf{Characteristic timescales}}\\
\noalign{\vskip 2pt}
$\tau_0$ & LLG damping / attempt time & s & $\tauD$ & Debye (rotational-diffusion) time & s \\
$\tauN$ & N\'eel relaxation time & s & $\tauB$ & Brownian relaxation time & s \\
$\tau_J$ & inertial decay time & s & $\tau_\mathrm{diff}$ & translational diffusion time & s \\
\noalign{\vskip 6pt}
\multicolumn{6}{@{}l}{\textbf{Mechanical and hydrodynamic quantities}}\\
\noalign{\vskip 2pt}
$\bv$ & particle velocity & m\,s$^{-1}$ & $\bomega$ & particle angular velocity & s$^{-1}$ \\
$\mathcal{M}$ & particle mass & kg & & & \\
$\vec{L}$ & total angular momentum & J\,s & $J$ & moment of inertia & kg\,m$^{2}$ \\
$\bF$ & force & N & $\bT^\mathrm{therm}$ & thermal stochastic torque & N\,m \\
$\vec{\Omega}$ & local fluid angular velocity & s$^{-1}$ & $\bUpsilon$ & hydrodynamic friction matrices & various \\
$\zeta^t$ & translational friction coefficient & kg\,s$^{-1}$ & $\zeta^r$ & rotational friction coefficient & kg\,m$^{2}$\,s$^{-1}$ \\
\noalign{\vskip 6pt}
\multicolumn{6}{@{}l}{\textbf{Geometry and numerical parameters}}\\
\noalign{\vskip 2pt}
$d$ & particle (core) diameter & m & $\dH$ & hydrodynamic diameter & m \\
$l$ & non-magnetic shell thickness & m & $d_\mathrm{sd}$ & critical single-domain diameter & m \\
$V_m$ & magnetic core volume & m$^{3}$ & $V$ & particle volume & m$^{3}$ \\
$a$ & equivalent hard-sphere radius & m & $r_{ij}$ & interparticle distance & m \\
$\Delta t$ & integration time step & s & $\Delta x$ & discretization cell size & m \\
\hline
\end{tabular}
\end{table*}

Let an MNP be subjected to a uniform magnetic field $\vec{H}_\text{ext}$ at a fixed temperature $T$. In this case, its total magnetic energy can be written as a sum of the Zeeman, demagnetization, exchange, and anisotropy contributions:
\begin{equation}
E = E_\text{ext} + E_\text{demag} + E_{\mathrm{ex}} + E_{\mathrm{ani}}.
\end{equation}
The Zeeman energy, $E_\text{ext}$, characterizes the interaction between the magnetization $\vec{M}$ and an externally applied magnetic field $\vec{H}_\text{ext}$ and is expressed as
\begin{equation}
E_\text{ext} = -\mu_0 \int_{V_m} \vec{M} \cdot \vec{H}_\text{ext} \, dV,
\label{eq:e_zeeman}
\end{equation}
where $V_m$ is the magnetic volume and $\mu_0$ is the magnetic permeability of vacuum. The Zeeman energy reaches its minimum when the magnetization points in the same direction as the external field. The demagnetization energy, $E_\text{demag}$, also commonly referred to as the stray-field energy, is given by the interaction of the magnetization with the demagnetization field it generates:
\begin{equation}
E_\text{demag} = -\frac{\mu_0}{2} \int_{V_m} \vec{M} \cdot \vec{H}_d \, dV,
\label{eq:e_demag}
\end{equation}
where $\vec{H}_d$ denotes the demagnetization field, which can be written more explicitly as a convolution of the magnetization with the demagnetization tensor $\widetilde{\bm{N}}$
\begin{equation}
\vec{H}_d(\br) = \int_{V_m} \widetilde{\bm{N}}(\br - \br') \, \vec{M}(\br') \, d\br',
\label{eq:hd_tensor}
\end{equation}
with the demagnetization tensor given by
\begin{equation}
\widetilde{\bm{N}}(\br - \br') = -\frac{1}{4\pi} \nabla \nabla' \frac{1}{|\br - \br'|}.
\end{equation}
In Eq.~\eqref{eq:hd_tensor}, $\br$ denotes the field point and $\br'$ the source point. $\nabla$ acts on $\br$ and $\nabla'$ on $\br'$. For the special case of uniformly magnetized spheres the demagnetization field is spatially uniform and can be expressed as $\vec{H}_d = -(1/3) \vec{M}$.

The exchange energy originates from the quantum-mechanical exchange interaction between neighboring spins. In the continuum approximation, it favors parallel alignment of nearby magnetic moments and penalizes spatial variations in the magnetization direction. It is given by
\begin{equation}
E_{\mathrm{ex}} = A \int_{V_m} (\nabla \bmag)^2 \, \mathrm{d}V,
\end{equation}
where $A$ is the exchange stiffness constant (in J/m) and $\bmag = \vec{M}/M_s$ is the reduced magnetization vector with $|\bmag| = 1$, and
\begin{equation}
 (\nabla \bmag)^2 = \sum_{i,j=x,y,z} \left( \frac{\partial m_i}{\partial j} \right)^2.
\end{equation}
Here, $M_s$ denotes material saturation magnetization. Finally, the magnetic anisotropy energy describes the preference of the magnetization to align along specific crystallographic directions, known as easy axes. This effect arises from spin–orbit coupling and the symmetry of the crystal lattice. For uniaxial anisotropy, the energy is commonly written as
\begin{equation} \label{eq:Eani}
E_{\mathrm{ani}} = K \int_{V_m} \left( 1 - (\bmag \cdot \baxis)^2 \right) \, \mathrm{d}V,
\end{equation}
where $K$ is the uniaxial anisotropy constant and $\baxis$ is a unit vector along the easy axis. The anisotropy energy is minimized when the magnetization aligns parallel (or antiparallel) to the easy axis.

Together, these energy contributions determine the static and dynamic magnetic properties of MNPs and whether they remain approximately uniformly magnetized (single-domain) or lower their energy through nonuniform magnetic structure (multidomain). While a single-domain state minimizes exchange and anisotropy energies, it generally produces a large demagnetization energy. The formation of magnetic domains reduces this demagnetization energy at the cost of introducing domain walls, which carry an associated wall energy. A simple criterion for the stability of the single-domain state can be obtained by comparing the demagnetization energy of a uniformly magnetized particle with the energy cost associated with domain-wall formation. For sufficiently small particles, the demagnetization energy is low enough that the system remains in a single-domain configuration. However, above a certain particle size, the reduction in demagnetization energy achieved by forming multiple domains outweighs the domain-wall energy penalty. For a spherical particle, this comparison leads to the concept of a \emph{critical single-domain diameter} $d_{\mathrm{sd}}$.
A commonly used estimate yields \cite{skomskimagnetism2008}
\begin{equation}\label{eq:macro_micro_rule}
 d_{\mathrm{sd}} \approx \frac{72 \sqrt{A K}}{\mu_0 M_s^2}.
\end{equation}
Particles with diameters $d \lesssim d_{\mathrm{sd}}$ have a single-domain ground state, while for $d \gtrsim d_{\mathrm{sd}}$ the energetically lowest state consists of multiple magnetic domains. Depending on material parameters, $d_{\mathrm{sd}}$ can range from a few tens of nanometers in soft magnetic materials to values on the order of micrometers in very hard magnets.

It is important to emphasize that in magnetic systems the existence of a lower-energy domain state does not necessarily imply that this state is realized in practice. Magnetic materials are well known to exhibit hysteresis, reflecting the fact that their magnetic state can depend strongly on the history of applied magnetic fields and prior magnetization processes. If a particle is prepared in a saturated single-domain state, for example by applying a strong external magnetic field, the transition toward a multidomain ground state requires the nucleation and propagation of domain walls. This process involves overcoming an energy barrier associated with exchange, anisotropy, and demagnetization contributions. As a result, the system may remain trapped in a metastable single-domain configuration for a long time even if this state is not the global energy minimum. The ability of a magnetic system to reach its ground state is therefore governed not only by energetic considerations but also by the height of the relevant energy barriers and the available observation or waiting time. If the energy barrier separating a metastable state from the true ground state is sufficiently large compared to thermal energy, the transition rate becomes negligibly small. In this case, the lowest-energy state cannot be reached within experimentally accessible timescales. This interplay between energetics, energy barriers, and magnetic history, well known from spin glasses \cite{lundgren_dynamics_1983,hiroi_superspin_2011}, underlies many technologically relevant phenomena, including magnetic hysteresis, coercivity, and thermal stability~\cite{kim2010biofunctionalized}.

If one adopts the homogeneously magnetized, single-domain MNP approximation, the particle is represented by a magnetic moment $\vec{\mu}=\mu\bmag$, with $\mu=M_s V_m$.

The Zeeman contribution \eqref{eq:e_zeeman} then simplifies to
\begin{equation}
u_H=-\mu_0\mu(\bmag\cdot \vec{H}_\text{ext}),
\label{eq:zee}
\end{equation}
and the corresponding field strength is conveniently measured by the Langevin parameter:
\begin{equation}
\xi= \frac{\mu_0\mu H_\text{ext}}{k_BT},
\label{eq:xi}
\end{equation}
that compares the dipole-field interaction energy to the thermal scale $k_B T$. In the same reduced single-domain picture, the uniaxial magnetic anisotropy energy Eq.~(\ref{eq:Eani}) could be rewritten (up to a physically inconsequential additive constant) as

\begin{equation} \label{eq:ani_sd}
  u_\text{ani} = - K V_m (\bmag \cdot \baxis)^2,
\end{equation}

\noindent
and the anisotropy barrier is measured by the dimensionless parameter
\begin{equation} \label{anisotropy-parameter}
 \sigma = \frac{KV_m}{k_B T}.
\end{equation}
Within the homogeneous single-domain approximation, the demagnetization energy in Eq.~\eqref{eq:e_demag} simplifies considerably. For two MNPs centered at $\br_i$ and $\br_j$, Eqs.~\eqref{eq:e_demag} and \eqref{eq:hd_tensor} contain self contributions, which describe the demagnetization energy of each individual particle, and a mutual contribution, which accounts for the interaction between the particles. Representing each particle by a single magnetic moment, $\vec{\mu}_i$ or $\vec{\mu}_j$, the latter reduces to the familiar dipole--dipole interaction,
\begin{equation}
{u_{ij}^\mathrm{dd}}= - \frac{\mu_0}{4\pi}
 \left[ 3\frac{\left(\vec{\mu}_i \cdot \br_{ij}\right)\left(\vec{\mu}_j \cdot \br_{ij}\right)}{r^5_{ij}} - \frac{\vec{\mu}_i \cdot \vec{\mu}_j}{r^3_{ij}} \right],
\label{eq:dipole-dipole}
\end{equation}
where $\br_{ij}=\br_i-\br_j$. The self-contributions are absorbed into the corresponding single-particle demagnetization energy, leaving Eq.~\eqref{eq:dipole-dipole} as the explicit interaction between particles. For equally sized particles with diameter $d$ and magnetic moment magnitude $\mu$, the interaction strength is conveniently characterized by
\begin{equation}
\lambda = \frac{\mu_0}{4\pi}\frac{\mu^2}{d^3 k_B T}.
\label{eq:lambda}
\end{equation}
This parameter provides a dimensionless measure of the strength of dipole--dipole interactions relative to thermal fluctuations, with $\lambda \sim 1$ indicating comparable dipolar and thermal energy scales.

In addition to determining the internal magnetic structure (single- or multidomain) of magnetic nanoparticles (MNPs), the interplay between the aforementioned energy scales also governs several magnetic relaxation mechanisms, each characterized by a distinct timescale. Note that the commonly used mathematical expressions for some of these timescales are derived within specific phenomenological models; therefore, the corresponding theoretical context is introduced together with each model. The relevant relaxation processes can be classified into three categories: internal magnetization dynamics, barrier-controlled relaxation of single-domain particles, and mechanical rotation of the particle as a whole.

When thermal fluctuations and external fields are absent, the shortest intrinsic magnetic relaxation timescale is the damping time of Larmor precession, $\tau_0$. From the standard Landau--Lifshitz--Gilbert (LLG) equation, it can be estimated as~\cite{poperechny2014dynamic}:
\begin{equation} \label{eq:tau0}
  \tau_0 = \frac{1}{\alpha \omega_L} = \frac{(1+\alpha^2)M_s}{2 \alpha \gamma K},
\end{equation}
where $\omega_L = \gamma \mu_0 H_\mathrm{ani}/(1 + \alpha^2)$ is the Larmor angular frequency in the absence of an applied field. Thus, $\tau_0$ characterizes the damping of the precession, whereas the precession period $2\pi/\omega_L$ characterizes the oscillatory motion. The phenomenological damping parameter $\alpha$ typically ranges between 0.01 and 0.1, and $\gamma \simeq 1.76 \times 10^{11}~\textrm{s}^{-1}\textrm{T}^{-1}$ is the gyromagnetic ratio.

At any finite temperature, the magnetic moment experiences thermal fluctuations. In the regime of weak external field and weak anisotropy, the typical timescale for its \emph{rotational diffusion}, known as the Debye time, is denoted by $\tauD$ and is given by:
\begin{equation}
\tauD = \frac{(1 + \alpha^2)\mu}{2 \alpha \gamma k_B T} = \sigma \tau_0.
\end{equation}
For MNPs with $\sigma \ge 1$, it is common to distinguish two mechanisms in the terminology of \citet{raikher2004nonlinear}: an intrawell relaxation, referring to fast fluctuations in the vicinity of a given energy minimum, and an interwell relaxation, referring to thermally activated jumps between the minima~\cite{raikher2004nonlinear,coffey_thermal_2012}. The interwell process is commonly known as N\'eel relaxation, with the characteristic timescale $\tauN$. In this context, $\tauD$ determines the prefactor of the interwell N\'eel time. The classical expression for the N\'eel time in~\citet{brown1963} is:
\begin{equation}\label{eq:tn_brown}
\tauN(\sigma \gg 1) = \frac{\tauD}{2\sigma}\sqrt{\frac{\pi}{\sigma}}\exp \sigma.
\end{equation}
If the particle is suspended in a viscous medium, two more relaxation times, connected to its mechanical rotation, must be introduced~\cite{raikher1994effective}. If a particle is coated with a non-magnetic shell of width $l$, its hydrodynamic diameter can be calculated as $\dH = d + 2l$. The inertial decay time is:
\begin{equation}
\tau_J = \frac{J}{\zeta^r} = \frac{\rho d^5}{60 \eta \dH^3},
\end{equation}
where $J = 0.1 \rho V_m d^2$ is the moment of inertia of an MNP, $\rho$ is the particle material density (here, we assume homogeneous spherical particles and neglect contributions from the steric shell),
$\zeta^r = \pi \eta \dH^3$ is the rotational friction coefficient, $\eta$ is the fluid viscosity.
\noindent
The second viscous relaxation time is the rotational Brownian time,
\begin{equation}
\tauB = \frac{\zeta^r}{2 k_B T} = \frac{\pi \eta \dH^3}{2 k_B T},
\end{equation}
the time needed for rotational diffusion of a spherical particle with diameter $\dH$ in a viscous solvent.

With this, we have outlined the relevant energy scales and timescales governing the dynamics of MNPs in viscous liquid carriers, deliberately excluding viscoelastic media effects, as these lie beyond the scope of the present work. In the following section, we use coupled translational and rotational Langevin equations as a representative particle-based framework for MNP motion and discuss how hydrodynamic interactions can be incorporated. This naturally extends into a broader discussion of how internal magnetization dynamics can be integrated into Molecular Dynamics simulations, thereby enabling scalable simulations of MNPs across different regimes.

\section{Modeling Translational and Rotational Diffusion of MNPs}\label{sec::md}

The dynamics of $N$ colloidal particles in a viscous, non-magnetic carrier fluid are governed by coupled translational and rotational Langevin equations~\cite{allenbook,Dhont_book}
\begin{align}
\mathcal{M}_j\dot{\bv}_j & = -\vec{\nabla}_j U + \bF_j^\mathrm{fric} + \bF_j^\mathrm{therm} \label{eq:maisF}\\
  \frac{\mathrm{d}}{\mathrm{d}t}\left(\bm{J}_j\bomega_j\right) & = -\vec{\mathcal{L}}_j U + \bT_j^\mathrm{fric} + \bT_j^\mathrm{therm} \label{eq:IdwisT}
\end{align}
with $j=1,\ldots,N$, and $\vec{\nabla}_j=\partial/\partial\br_j$ and $\vec{\mathcal{L}}_j=\bmag_j\times\partial/\partial\bmag_j$ the gradient and rotational operator with respect to particle $j$, respectively.
Here, $U$ is the potential energy, $\mathcal{M}_j$ is the mass of particle $j$, and $\bm{J}_j$ is its moment-of-inertia tensor. The derivative $\mathrm{d}(\bm{J}_j\bomega_j)/\mathrm{d}t$ is the inertial-frame time derivative of the particle's angular momentum. In body-fixed coordinates, it contains both $\bm{J}_j\dot{\bomega}_j$ and the gyroscopic term $\bomega_j\times(\bm{J}_j\bomega_j)$.
The first, second, and third terms on the right-hand side of Eq.~\eqref{eq:maisF} represent the systematic, friction, and fluctuating forces, respectively, and correspondingly for the torques in Eq.~\eqref{eq:IdwisT}.
Equations \eqref{eq:maisF} and \eqref{eq:IdwisT} can be thought of as Newton and Euler equations for linear and rotational motion, respectively, with additional terms to account for solvent effects at the single particle level.

In the simplest and most commonly used free-draining approximation, hydrodynamic interactions are neglected. The translational and rotational motions of MNPs through viscous solvents are opposed by friction forces and torques that depend only on the respective particle velocities~\cite{Morimoto2002,allenbook,ilg_magnetoviscosity_2005}
\begin{equation} \label{FfricTfric}
\bF_j^\mathrm{fric} = -\zeta^t_j\, \bv_j^\mathrm{p}, \quad
\bT_j^\mathrm{fric} = -\zeta^r_j\, \bomega_j^\mathrm{p},
\end{equation}
with no translation-rotation coupling.
Note that $\bv_j^\mathrm{p}$ denotes the translational velocity of MNP $j$ relative to the local velocity of the fluid flow.
Similarly, $\bomega_j^\mathrm{p}$ should be interpreted as the angular velocity relative to the local fluid angular velocity, \textit{i.e.}, half the local vorticity of the flow.
For hard spheres with no-slip boundary conditions, the translational and rotational friction coefficients are given by
\begin{equation} \label{eq:spherical_friction}
\zeta^t_j=6\pi\eta a_j, \qquad \zeta^r_j=8\pi\eta a_j^3,
\end{equation}
where $\eta$ denotes the dynamic viscosity of the solvent and $a_j\approx \dH/2$ the equivalent hard-sphere radius of MNP $j$. The general tensorial friction formulation for rigid bodies is introduced below in the discussion of hydrodynamic interactions.

The fluctuation forces and torques in Eqs.~\eqref{eq:maisF}, \eqref{eq:IdwisT} are modeled as random variables with zero mean.
The fluctuation-dissipation theorem relates the variance of these random variables to the friction coefficients.
In the free-draining approximation, these are
\begin{align}\label{eqn:fluct_diss}
\langle \bF_j^\mathrm{therm}(t) \otimes \bF_k^\mathrm{therm}(t') \rangle &= 2k_\mathrm{B}T \, \zeta^t_j \, \delta_{jk}\,\bm{I}\delta(t-t'),\\
\langle \bT_j^\mathrm{therm}(t) \otimes \bT_k^\mathrm{therm}(t') \rangle &= 2k_\mathrm{B}T \, \zeta^r_j \, \delta_{jk}\,\bm{I}\delta(t-t'),
\end{align}
with $\bm{I}$ the three-dimensional unit matrix,
whereas the remaining correlations vanish. Equations \eqref{eq:maisF} and \eqref{eq:IdwisT} lend themselves particularly well for simulations. There, the systematic potential terms $\vec{\nabla}_j U$ and $\vec{\mathcal{L}}_j U$ are typically determined by external fields and by interparticle interactions, for example from dipolar and soft-sphere interactions.
The latter provides the excluded volume interaction, preventing particles from overlapping by a significant amount. The term `soft` refers to the fact that the potential is continuous and differentiable with respect to distance, preventing the occurrence of infinite forces when integrating the equation of motion. A typical choice is the Lennard-Jones potential:
\begin{equation}
  u^\mathrm{LJ}_{ij} = 4\varepsilon\left[\left(\frac{d}{r_{ij}}\right)^{12}-\left(\frac{d}{r_{ij}}\right)^{6}\right],
  \label{eq:LJ}
\end{equation}
where $\varepsilon$ is the attractive well depth, $d$ is the particle diameter, and $r_{ij}=|\br_i-\br_j|$, as previously mentioned, denotes the pair distance between particles $i$ and $j$; or its purely repulsive variant, the Weeks-Chandler-Andersen potential~\cite{weeks1971role}:
\begin{equation}
  u^\mathrm{WCA}_{ij}=
\begin{cases}
u_\mathrm{LJ}(r_{ij})-u_\mathrm{LJ}(r_\mathrm{cut}),& \mathrm{if~} r_{ij}<r_\mathrm{cut}\\
 0 & \mathrm{otherwise}
\end{cases},
\label{eq:wca}
\end{equation}
\noindent
where $r_\mathrm{cut} = 2^{1/6}d$.
Typically, the length scale $d$ is set to the core diameter of the MNP, and the energy scale $\varepsilon$ is set to the thermal energy $k_BT$.

\subsection{Including Hydrodynamic Interactions}\label{subsec::hydrodynamics}

Hydrodynamic effects become important in a variety of situations. This is typically the case in dense systems, such as MNP suspensions embedded in polymers or gels, where flow fields generated by neighboring particles cannot be neglected. They also play a significant role when an external solvent flow is imposed, for example through shear. Furthermore, hydrodynamic interactions become increasingly relevant for strongly aspherical particles, actively propelled MNPs, or particles moving in close proximity to confining boundaries. The Langevin equation in the free-draining approximation only brings in solvent effects at the single-particle level. Taking hydrodynamics into account removes this limitation: if one MNP moves or rotates, this creates a fluid flow which, in turn, is felt by other nearby particles. Hence, the translational and rotational motions of nearby particles become correlated through the solvent flow. This hydrodynamic coupling is a well-known phenomenon in colloidal science and is not only present between MNPs, but also between MNPs and other components of the system, \textit{e.g.}, a polymer.

In general, $\vec{F}_j^\mathrm{fric}$ and $\vec{T}_j^\mathrm{fric}$ in Eqs.~\eqref{eq:maisF} and \eqref{eq:IdwisT} depend on the translational and rotational velocities of all particles.
The fast propagation of fluid disturbances relative to particle motion allows us to linearize the hydrodynamic equations, leading to the following relations~\cite{cichocki_friction_1994}:
\begin{align}
\vec{F}_j^\mathrm{fric} & = -\sum_k \boldsymbol{\Upsilon}_{jk}^\mathrm{TT}\cdot\vec{v}_k - \sum_k \boldsymbol{\Upsilon}_{jk}^\mathrm{TR}\cdot\vec{\omega}_k \ , \\
\vec{T}_j^\mathrm{fric} & = -\sum_k \boldsymbol{\Upsilon}_{jk}^\mathrm{RT}\cdot\vec{v}_k - \sum_k \boldsymbol{\Upsilon}_{jk}^\mathrm{RR}\cdot\vec{\omega}_k \ .
\end{align}
In practice, the friction matrices $\boldsymbol{\Upsilon}$ can be obtained to an excellent approximation from the solution to Stokes flow, where the particle-level Reynolds number is low.
The free-draining approximation introduced above corresponds to the limiting case in which the friction matrices are diagonal and the translational–rotational hydrodynamic coupling vanishes:
\begin{align}
 \boldsymbol{\Upsilon}_{jk}^\mathrm{TT} &\approx \zeta^t_j\, \bm{I} \delta_{jk}, \\
\boldsymbol{\Upsilon}_{jk}^\mathrm{RR} &\approx \zeta^r_j\, \bm{I} \delta_{jk}, \\
\boldsymbol{\Upsilon}_{jk}^\mathrm{TR} &\approx \boldsymbol{\Upsilon}_{jk}^\mathrm{RT} \approx 0.
\end{align}

Accurately accounting for hydrodynamic interactions can require substantial computational effort, especially when one aims to resolve the hydrodynamic influence on particle rotation \cite{Satoh99}.
In simulations, we typically include hydrodynamic interactions by coupling the equations of motion of the MNP to a solver for the Navier-Stokes hydrodynamic equations.
Several algorithms are available to solve the latter. Common choices for soft matter systems are the Lattice-Boltzmann method~\cite{mcnamara1988use,kruger2017lattice,kim09c} and Multi-Particle Collision Dynamics~\cite{gompper09,kapral2008multiparticle,zablotsky_field_2019}.
Here, we focus on the Lattice-Boltzmann algorithm. It is fast, can handle thermal fluctuations~\cite{duenweg07a} and complex geometries, is well suited for parallel computing and GPUs, and is available in open-source simulation codes.

Two levels of coupling between particle and solvent dynamics are commonly used in simulations: either just at the center of mass of the particle, or for the entire particle volume or surface. Coupling restricted to the particle centers of mass, often referred to as point coupling, connects only the translational motion of the particles with the fluid.
In this case, one can modify the translational Langevin equation under the free-draining approximation by introducing a two-way coupling between the motion of the MNPs and the solvent~\cite{ahlrichs99a}:
\begin{equation}
\mathcal{M}_j\dot{\vec{v}}_j = -\vec{\nabla}_j U + \vec{F}_j^\mathrm{coupling},
\end{equation}
where the coupling force comprises dissipative and stochastic contributions. In the point-coupling scheme, the friction coefficient entering both contributions is a bare particle--fluid coupling parameter $\zeta_j^0$. The coupling force is
\begin{equation}
  \vec{F}_j^\mathrm{coupling} = -\zeta_j^0 \left[\vec{v}_j - \vec{V}^{\mathrm{fluid}}(\vec{r}_j)\right] + \vec{F}_j^\mathrm{therm,LB},
\end{equation}
The stochastic contribution has zero mean and retains the translational fluctuation--dissipation relation in Eq.~\eqref{eqn:fluct_diss}, with the free-draining coefficient $\zeta_j^t$ replaced by the bare coupling coefficient $\zeta_j^0$. The lattice-Boltzmann fluid degrees of freedom must likewise be thermalized consistently~\cite{duenweg07a}. Because the interpolated fluid responds to the coupling force, $\zeta_j^0$ should not in general be identified with the isolated-particle Stokes coefficient $\zeta_j^t$ in Eq.~\eqref{eq:spherical_friction}. The effective hydrodynamic friction also depends on the fluid viscosity, lattice spacing, and interpolation scheme, and can be calibrated to reproduce the desired hydrodynamic radius~\cite{ahlrichs99a,fischer15}.

To conserve total momentum, the equal and opposite complete coupling force, including its dissipative and stochastic contributions, is distributed to the surrounding lattice-Boltzmann nodes, using the same interpolation weights employed to obtain the local fluid velocity. In this way, the particle velocity and the local fluid velocity tend to align.

These two couplings affect only translations. The rotational dynamics is still handled by the rotational Langevin equation in the free-draining limit. Moreover, coupling at a single point for each particle implies that hydrodynamic effects of the particle shape are neglected, and the flow field close to the particle will not be accurately resolved.
The main advantage of the point-coupling scheme is the relatively low computational effort: typically, the grid spacing of the lattice-Boltzmann solver is set to the particle diameter, giving manageable grid sizes and computations on the order of milliseconds per time step for a typical ferrofluid simulation with $N=1000$ particles.
This coupling scheme has been used when studying long-ranged couplings introduced by hydrodynamics~\cite{novikau26}.

A more precise coupling is then achieved by imposing no-slip boundary conditions for the fluid at the MNP surface. In other words, we require that the fluid velocity at any point on the MNP surface is equal to the motion of the particle surface, both from particle translation and rotation.
There are two ways to achieve this. First, one can directly impose the boundary condition onto the fluid, keeping track of the momentum change that is needed to fulfil it. The momentum transfer integrated over the entire particle surface is then applied as force and torque onto the particle~\cite{ladd94am,kim09c}. Although this approach most closely matches the mathematical intent, it demands substantial modifications to the simulation code.

\begin{figure}[h!]
\centering
\includegraphics[width=\linewidth]{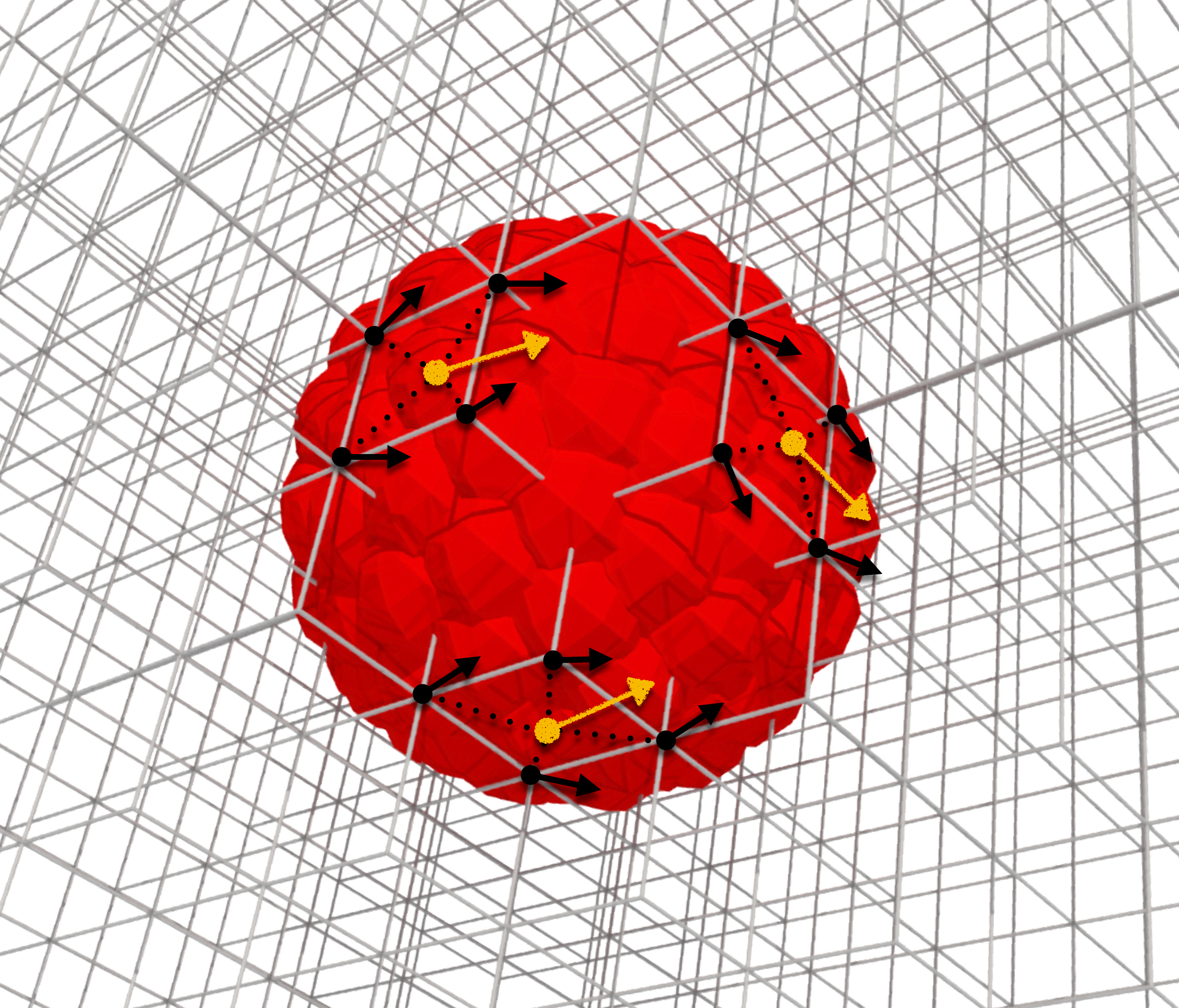}
\caption{Simulation rendering of a raspberry MNP embedded in a lattice-Boltzmann grid, illustrating the frictional coupling between the fluid and the particles. The local fluid velocity vectors are shown as black arrows. These velocities are trilinearly interpolated from the lattice nodes to the positions of the particles within the raspberry MNP (yellow arrow). A frictional force is then applied to each particle based on its velocity relative to the interpolated local fluid velocity.}
\label{fig:raspberry}
\end{figure}

A commonly used alternative is the so-called raspberry model~\cite{lobaskin04a,fischer15}. Here, one models the MNP as an assembly of rigidly connected coupling points, so-called virtual sites~\cite{arnold13a}, which follow the MNP translation and rotation. These coupling points are then all coupled to the LB fluid via the point-coupling scheme described above. A visual depiction of a raspberry MNP in an LB fluid, with the frictional coupling scheme, is provided in Fig.~\ref{fig:raspberry}. Again, the forces on the coupling points are accumulated into a force and torque applied to the MNP. If the coupling points are placed densely throughout the MNP volume, rotational and translational drag coefficients consistent with those for spheres and ellipsoids in Newtonian fluids can be reproduced. However, the process for constructing the raspberries is not obvious and somewhat empirical. All of the aforementioned approaches provide accurate coupling of rotational and translational motion to the hydrodynamics solver, at the cost of a significantly higher computational effort: to capture the shape of the particle well enough, despite the cubic lattice used by the lattice-Boltzmann solver, approximately 8 grid points are needed per particle diameter.

\section{Magnetodynamics in Colloidal Simulations}\label{sec::mmodels}

Having summarized the relevant energy and timescales involved in the phenomenology of MNPs in Section~\ref{sec::relevant_scales}, in this section, we discuss the available analytical and numerical approaches that can be combined with computational schemes such as those presented in Section~\ref{sec::md} to explicitly include magnetodynamics in colloidal MNP simulations. Here, we use \emph{magnetodynamics} as an umbrella term for the magnetic models considered, whereas \emph{magnetization dynamics} refers specifically to the temporal evolution of the particle magnetization. The section is organized like a decision tree that an aspiring researcher interested in colloidal MNP simulations can follow to find the optimal magnetic level of description required for their specific problem. The primary question to answer is whether the MNPs can be treated as homogeneously magnetized and single-domain (using a criterion such as Eq.~\ref{eq:macro_micro_rule}), based on the material parameters. If the MNP diameter is larger than the \emph{critical single-domain diameter} $d_{\mathrm{sd}}$, there is not much flexibility, and unless there are some specific mitigating circumstances (\textit{i.e.,} sample preparation), a micromagnetics approach is warranted, as described in Section~\ref{subsec:mmodel_multi}. On the other hand, if the diameter of the MNPs of interest is smaller than $d_{\mathrm{sd}}$, there are quite a few methods available, as described in Section~\ref{subsec:mmodel_single}. There, one can make various compromises between computational cost and accuracy, depending on the anisotropy parameter of the MNPs of interest and, related to that, the smallest timescale that needs to be resolved for the phenomenology of interest. Before we proceed to the model discussion proper, it is worth mentioning that a number of phenomenological models have been proposed in the literature based on nonequilibrium thermodynamics~\cite{felderhof_hydrodynamics_1999,muller_structure_2001} that are beyond the scope of this article.

\subsection{Single-domain Particles}\label{subsec:mmodel_single}

\begin{figure*}[h!]
\centering
\includegraphics[width=0.8\linewidth]{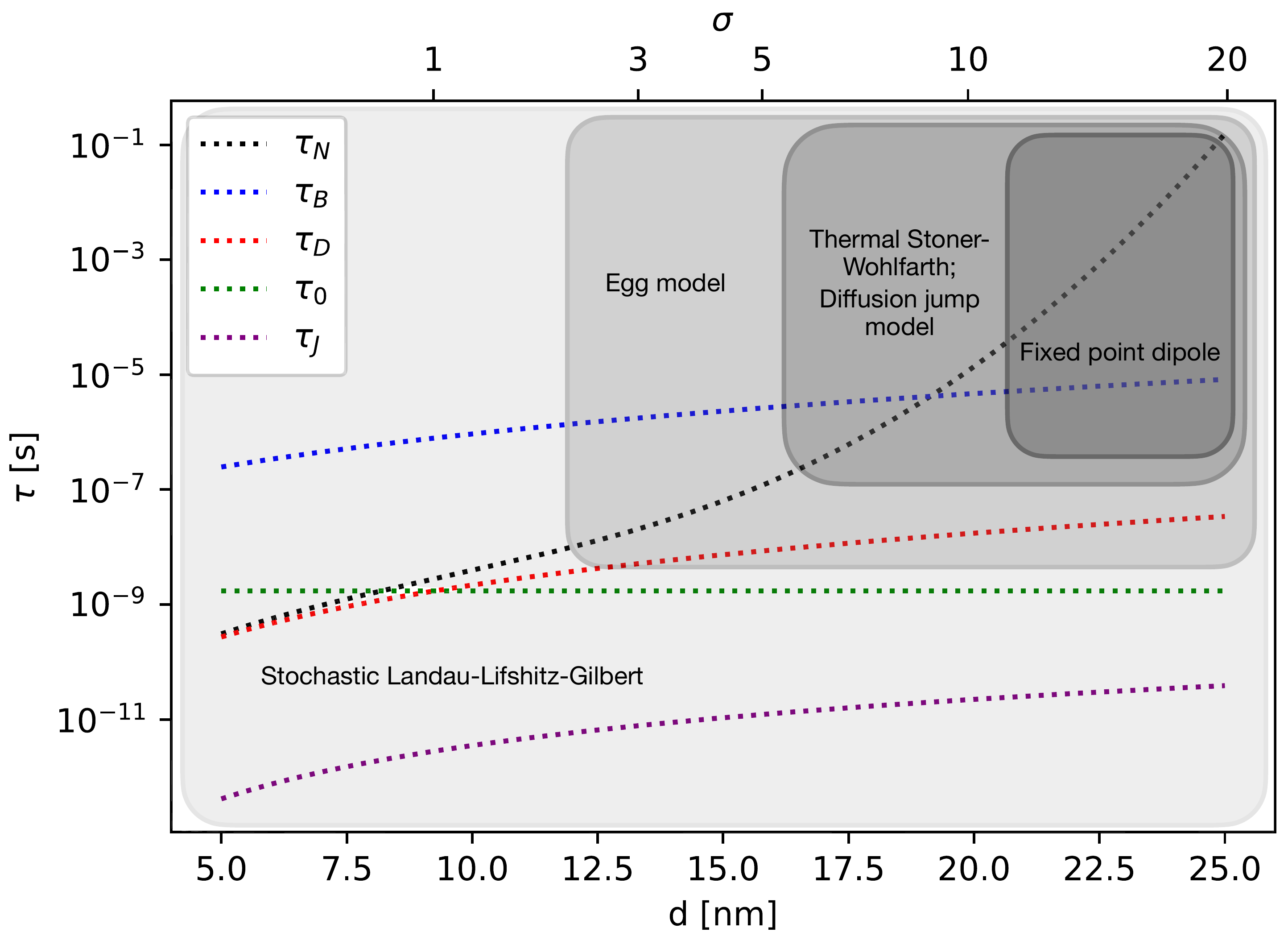}
\caption{Characteristic timescales of the relaxation processes for MNP ensembles. Timescales are calculated for magnetite ($M_s=480\,\mathrm{kA\,m^{-1}}$, $K=10\,\mathrm{kJ\,m^{-3}}$, $\rho=5170\,\mathrm{kg\,m^{-3}}$, $\alpha=0.08$, $T=298\,\mathrm{K}$). The MNPs have a uniform nonmagnetic shell of radial thickness $2\,\mathrm{nm}$. The MNPs are suspended in water ($\eta=0.89\,\mathrm{mPa\,s}$). The top x-axis shows $\sigma$ corresponding to the particle size on the bottom x-axis at the specified temperature. Colored regions indicate the approximate applicability regimes of the corresponding magnetodynamic models in terms of particle size, characteristic timescale, and $\sigma$. These regimes may overlap, and additional restrictions, including those associated with interaction strength, are specified in the text. The relative area of each box qualitatively represents its computational cost.}
\label{fig:perspectives_overview_doodle}
\end{figure*}

Figure~\ref{fig:perspectives_overview_doodle} complements this discussion with a visual overview. The recommendations are intended as general guidelines rather than universally applicable prescriptions. Model selection ultimately requires a more detailed, problem-specific assessment based on the broader considerations developed throughout this Perspective.
As a representative example, we consider magnetite MNPs suspended in water and coated with a \(2~\mathrm{nm}\)-thick oleic acid layer. Beginning at the critical single-domain diameter \(d_{\mathrm{sd}}\) defined by Eq.~\eqref{eq:macro_micro_rule}, we consider progressively smaller particles. Figure~\ref{fig:perspectives_overview_doodle} shows the characteristic relaxation times (dotted lines), together with colored regions indicating the approximate regimes of applicability of the models considered.

Note that magnetite has a cubic anisotropy~\cite{witt2005three}, and the \(K\) value used here is an effective uniaxial anisotropy consistent with experimental estimates for isolated magnetite nanoparticles. The effective barrier can be inferred from the blocking condition for a given attempt frequency, \(\tau_{\mathrm N}(T_{\mathrm B})\sim t_{\mathrm m}\), where $T_B$ is the blocking temperature and $t_m$ is the measurement time. \citet{mamiya2020estimation} obtained effective anisotropy distributions in the range \(10\)--\(20~\mathrm{kJ/m^3}\) and observed a remanence-to-saturation magnetization ratio of 0.5, consistent with predominantly uniaxial-like reversal in a non-interacting system. Using an effective uniaxial anisotropy is a common and practical strategy that should be understood as a barrier-equivalent parameter incorporating the combined influence of magnetocrystalline, surface, shape, and defect contributions. The same effective barrier can then be used directly in simulations to reproduce the corresponding Néel relaxation time within the adopted thermal-activation model. This provides a self-consistent reference parameter set for comparing the characteristic timescales and predictions of different magnetodynamic models without requiring each model to resolve the full microscopic origin or symmetry of the anisotropy energy. Because of this, our discussion of single-domain MNPs is formulated strictly in terms of uniaxial anisotropy. Incorporating other forms of anisotropy is possible in many of the models discussed, although the practical complexity is model dependent, and more information can be found in the relevant references provided for each model. Our focus here remains on a comprehensive overview and comparison of the available models. Note that the relative performance characteristics between models are provided based on the effective uniaxial anisotropy approach and may differ for other types of anisotropy.

As the particle diameter changes, so does its anisotropy energy. The relevant magnetic relaxation mechanisms are determined by the ratio between the anisotropy and thermal energies. Therefore, the anisotropy parameter $\sigma$ is the key parameter to consider. If the MNPs have a high magnetic anisotropy $\sigma \gtrsim 15$, this regime corresponds to the smallest, darkest rectangle on the right side of Fig.\ \ref{fig:perspectives_overview_doodle}. For this case ($\tauB\ll \tauN$), Brownian relaxation is the dominant magnetic relaxation mechanism, and the point-dipole approximation discussed in Section~\ref{subsec:fpd} is valid. The effective field method described in Section~\ref{subsec::effld} shares this timescale regime but additionally assumes a dilute or effectively noninteracting suspension.

For MNPs in the $5<\sigma<15$ range, the only two relevant relaxation mechanisms to consider are the Brownian and N\'eel relaxation. In the case of $\tauB \sim \tauN$, the thermal Stoner-Wohlfarth (tSW) model, discussed in Section~\ref{subsec:sw}, and the Diffusion-Jump model, discussed in Section~\ref{subsec:DJ}, are well-suited approaches. However, it is hard to make sweeping recommendations because additional relaxation mechanisms can become relevant depending on the material or system specificities. As a rule of thumb, we do not recommend using the diffusion-jump model for $\sigma<8$, and we do not recommend using the tSW model for $\sigma<5$.

For $\sigma < 5$, when the N\'eel (``inter-well'') relaxation timescale becomes comparable to that of the ``intra-well'' dynamics, but one still aims to avoid explicitly resolving the inertial timescales, which are typically on the order of picoseconds, the Egg model described in Section~\ref{subsec:egg} is the appropriate framework to use.

Naturally, assuming unlimited computational power and time, the obvious recommendation is to employ the macrospin Landau–Lifshitz–Gilbert equation, presented in Section~\ref{subsec:macrospin_llg}, since it constitutes the most accurate model currently available.

\subsubsection{Fixed-Point Dipole}\label{subsec:fpd}

In the fixed-point dipole (thermally blocked) regime, $\tauB \ll \tauN$, the magnetic moment is rigidly locked to the particle body (rigid-dipole approximation), so that the accessible degrees of freedom reduce to the particle positions and dipole orientations, $\{\br_1,\bmag_1,\ldots,\br_N,\bmag_N\}$. Within the transport framework introduced in Section \ref{sec::md}, this is the homogeneous single-domain frozen-moment specialization in which the particle-body rotation and the magnetic-moment rotation are identified.

The above model, using the free-draining approximation, has been used frequently in the literature to simulate, \textit{e.g.}, the magnetization and aggregation dynamics, as well as magnetoviscous effects~\cite{ilg_magnetoviscosity_2005,Ilg_lnp,cerda_aggregate_2008,jordanovic_structure_2009,Sreekumari2013,Sreekumari2015,rosa_shear_2020,zverev_computer_2021}. 
In fact, progress in the field has been mainly driven by such many-body computer simulations with additional insights from dynamic mean-field theories~\cite{ilg_magnetoviscosity_2005,camp_how_2021,ivanov_effects_2022}.
While the model simulates the dynamics of MNPs, 
it is also useful to obtain equilibrium properties of systems containing MNPs. It has been used to study the structure of standard ferrofluids~\cite{wang02a,cerda_aggregate_2008}, but also cluster formation for more complex MNPs with different shapes and internal structures~\cite{2013-kantorovich-sm,steinbach16a,rosenberg2023influence}, as well as for MNPs embedded in polymer gels~\cite{weeber15}. The model can also be used to study MNP dynamics, provided that the dipole moments relax solely through Brownian rotation and that hydrodynamic coupling can be neglected. Simulations based on this model are straightforward to run. Simulation codes such as ESPResSo~\cite{weik19a,grad_2026_20450775}, LAMMPS~\cite{thompson22a}, and InnMNP~\cite{InnMNP2026} have the required components built in, namely, the solution of the translational and rotational Langevin equations, various soft-sphere potentials, and solvers for dipolar interactions under open, periodic~\cite{2008-cerda-jcp}, and mixed~\cite{brodka04a,weeber19b} boundary conditions.

When hydrodynamic couplings are important, the transport extensions introduced in Section \ref{sec::md} can be combined with this fixed-point-dipole model. Point coupling has been used to study long-ranged solvent-mediated effects~\cite{novikau26}, while more resolved moving-boundary and raspberry couplings have been used for ferrofluid dynamics and AC-susceptibility calculations~\cite{kim09c,kreissl21,kreissl23}.

\paragraph{Limits and Performance Considerations}

As already suggested, the model is only applicable when the relaxation of the orientation of the magnetic moment of an individual MNP can be described by a single timescale, which is chosen by setting the friction coefficient in the rotational Langevin equation. The presence of the inertia terms in Eqs.~\eqref{eq:maisF}, \eqref{eq:IdwisT} enforces a very short reference time, which, in practice, limits the available timescales that can reasonably be covered in the simulations. This limitation is more severe for interacting simulations due to the additional computational costs. To extend the simulation time window, the overdamped limit $\mathcal{M}_j\dot{\bv}_j\to 0, \frac{\mathrm{d}}{\mathrm{d}t}\left(\bm{J}_j\bomega_j\right)\to 0$ is often invoked in colloidal science \cite{Dhont_book}. This assumption turns Eqs.~\eqref{eq:maisF} and \eqref{eq:IdwisT} into force and torque balance equations, respectively, often called Brownian dynamics to distinguish them from Langevin dynamics, which includes inertia terms. In the context of MNPs and ferrofluids, relatively few studies have made use of the overdamped limit for free-draining interacting systems (see \textit{e.g.}~\cite{pi_hex,ilg_structure_2006}).

\subsubsection{Effective Field Method}\label{subsec::effld}

For a suspension of thermally blocked MNPs,
the so-called effective field method can be viewed as a viable alternative to a direct numerical integration of 
Eq.\ \eqref{eq:IdwisT} in cases when interparticle interactions can be neglected. The method was originally developed to study the orientational kinetics of ferrofluids~\cite{martsenyuk1974kinetics,raikher1994effective} and was recently successfully adopted in a number of superparamagnetic systems, including viscoelastic ferrocolloids~\cite{Rusakov2021}, magneto-induced flows in thrombosed channels~\cite{MUSIKHIN2023}, and magnetically-steered chiral nanobots~\cite{Tripathi2025}. The main assumption is that the probability density for the magnetic moment to have a specific orientation $\bmag$ at time $t$ can always be represented by a Boltzmann-like distribution
\begin{equation}
  W(\bmag,t) = \frac{\exp\Big(\bxi_\mathrm{eff}(t) \cdot \bmag\Big)}{\int \exp\Big(\bxi_\mathrm{eff}(t) \cdot \bmag\Big) d \bmag} , \label{eq:wefa}
\end{equation}
\noindent
where $\bxi_\mathrm{eff}$~is a time-dependent dimensionless {effective field}.
In thermodynamic equilibrium, the latter coincides with the external magnetic field $\bxi = \mu_0 \mu \vec{H}_\mathrm{ext}/k_B T$.
However, in any non-equilibrium scenario, $\bxi_\mathrm{eff}$ has to be determined
self-consistently by substituting the ansatz~(\ref{eq:wefa}) into the Fokker-Planck equation corresponding to Eq.~\eqref{eq:IdwisT} in the overdamped limit and in the absence of interparticle interactions.
For a dilute suspension of spherical MNPs subjected to a hydrodynamic flow,
the effective field method produces the following equation
for the ensemble-averaged magnetic moment $\eav = \int \bmag W(\bmag,t)d \bmag$~\cite{Shliomis2001}:

\begin{multline}\label{eq:mrsh}
  \frac{d}{dt} \eav = \vec{\Omega} \times \eav - \left[ 1 - \frac{\bxi \cdot \bxi_\mathrm{eff}}{\xi^2_\mathrm{eff}}\right]\frac{\eav}{\tauB} \\ - \left[ 1 - \frac{L(\xi_\mathrm{eff})}{\xi_\mathrm{eff}} \right] \frac{\eav \times \Big(\eav \times \bxi\Big)}{2 \tauB L^2(\xi_\mathrm{eff})},
\end{multline}

\noindent
where $\vec{\Omega} = \frac{1}{2} \bnabla \times \vec{V}^\mathrm{fluid}$ is the local fluid angular velocity, or half the local flow vorticity, and $L(\xi) = \coth \xi - 1/\xi$ is the Langevin function.
To close Eq.~\eqref{eq:mrsh}, one 
{uses Eq.\ \eqref{eq:wefa} to find the}
non-linear relationship $\eav = L(\xi_\mathrm{eff})\bxi_\mathrm{eff}/\xi_\mathrm{eff}$.
For practical evaluations, it might be more convenient to exclude the effective field through $\bxi_\mathrm{eff} = L^\text{inv}(|\eav|)\eav/|\eav|$, where $L^\text{inv}$ is the inverse Langevin function~\cite{BENITEZ2018}.
For weakly non-equilibrium problems, \textit{i.e.}, for $|\bxi_\mathrm{eff} - \bxi | \ll \xi$, Eq.~(\ref{eq:mrsh}) can be simplified through linearization~\cite{raikher1994effective}.

\paragraph{Limits and Performance Considerations}

It is important to emphasize that Eq.~(\ref{eq:mrsh}) is an {approximation} for the true dynamics
of thermally blocked MNPs described by 
Eq.\ (\ref{eq:IdwisT}) in the non-interacting limit.
However, direct comparisons with Langevin dynamics simulations demonstrated its accuracy even in strongly non-linear problems far from equilibrium~\cite{BlumsCebersMaiorov1997,Rinaldi2010,Kuznetsov2022}. Unlike Eq.\ (\ref{eq:IdwisT}), Eq.~(\ref{eq:mrsh}) is deterministic rather than stochastic; therefore, its numerical solution does not require gathering sufficient statistics, making it much more computationally efficient.
Moreover, Eq.~(\ref{eq:mrsh}) can be coupled with other continuous-level equations ({\it e.g.}, Navier-Stokes, advection-diffusion, and Maxwell's) to account for nonequilibrium magnetization dynamics on a macroscopic scale~\cite{Krekhov2017,PhysRevE.106.064605}.

\subsubsection{Thermal Stoner--Wohlfarth}\label{subsec:sw}

\begin{figure}
  \centering
  \includegraphics[width=0.7\linewidth]{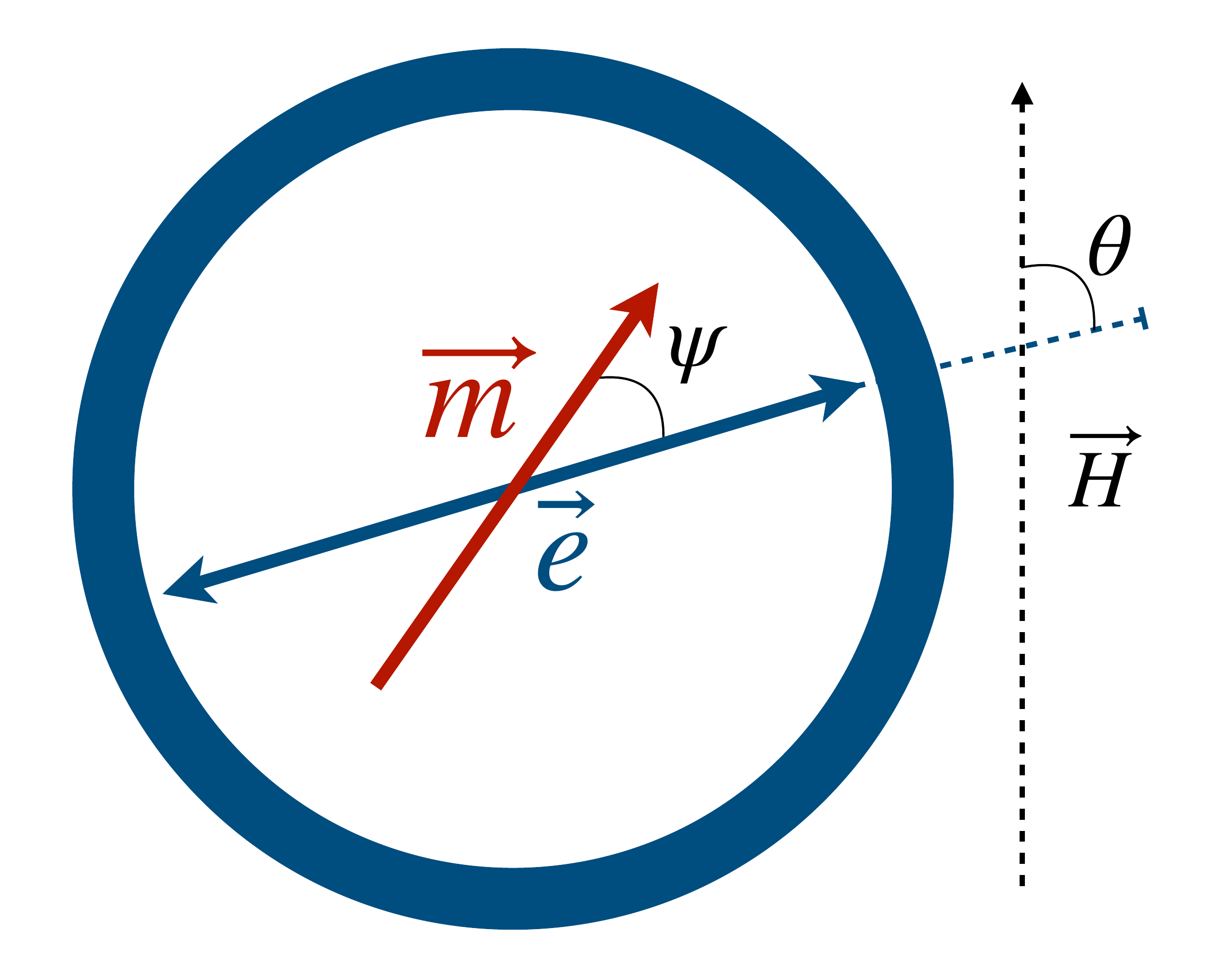}
  \caption{Geometric details and relevant angles used in the formulation of the total magnetic energy in the Stoner-Wohlfarth model. $\theta$ is the angle between $\vec{e}$ and $\vec{H}_\text{ext}$, and $\psi$ is the angle between $\bmag$ and $\vec{e}$.}
  \label{fig:sw_angles}
\end{figure}

To explain magnetic hysteresis effects in alloys, Stoner and Wohlfarth (SW) introduced the concept of single-domain ``particles'' for spatial domains small enough that magnetic domain boundaries are energetically unfavorable~\cite{stoner_mechanism_1948} when considering Zeeman \eqref{eq:zee} and magnetic anisotropy \eqref{eq:ani_sd} energies,
\begin{equation}\label{eq:sw_mag_ene}
  U = - \mu_0 \mu(\bmag\cdot\vec{H}_\text{ext}) - KV_m (\bmag \cdot \baxis)^2 .
\end{equation}
In this classical model, the equilibrium configuration at zero temperature is determined by the balance of Zeeman and magnetic anisotropy energy of such particles. Rather intuitively, resolving the dynamics of the dipole moment in this model can be understood as finding the angle between $\vec{e}$ and $\vec{m}$, $\psi$, for a given angle between $\vec{e}$ and $\vec{H}$, $\theta$, that minimizes Eq.~\ref{eq:sw_mag_ene}. A visual depiction of the classical Stoner--Wohlfarth particle is provided in Fig.~\ref{fig:sw_angles}. More recently, fundamental inequalities within the SW model were derived that bound the initial magnetic susceptibility and determine a borderline value of the uniaxial anisotropy constant~\cite{iglesias_fundamental_2022}. It is tempting to transfer the SW model to MNPs with magnetic cores small enough so that they can be considered single-domain particles. However, the zero temperature assumption in the SW model and the corresponding absence of thermal fluctuations is a severe and rather unrealistic approximation when applied to MNP systems.

To better describe hysteresis loops of single-domain MNPs, several authors have suggested extending the classical Stoner-Wohlfarth model by including thermal fluctuation-induced transitions between the two minima of the Stoner-Wohlfarth energy function~\cite{lu_field_1994,chuev_nanomagnetism_2007,usov_hysteresis_2009,carrey2011simple}. In particular, Kramers-type rate equations for the population of energy minima have been proposed. These approaches can therefore be seen as combining the SW model with N\'eel's picture of magnetization relaxation by thermally activated magnetization reversals~\cite{chuev_nanomagnetism_2007} along the lines of the classical work by Bean and Livingston~\cite{bean_superparamagnetism_1959}. These extensions to the classical Stoner-Wohlfarth model are sometimes referred to in the literature as the generalized SW model~\cite{chuev_nanomagnetism_2007}. The thermal SW (tSW) is an implementation of the generalized SW model that strictly adheres to the two-state approximation~\cite{mostarac2025thermal,wolfschwenger2024molecular}. In this approach, the continuous magnetization trajectory is replaced by thermally activated transitions between energy minima separated by field-dependent energy barriers. The field-dependent energy barriers are computationally straightforward and fast to estimate from the extrema of Eq.~\eqref{eq:sw_mag_ene}. Assuming that the processes are thermally activated, transition-state theory can be applied to determine, from the energy barriers, the probability for the dipole moment to flip its orientation, thus reproducing Néel relaxation. When reverse switching within a time step is neglected and flips from the currently occupied minimum are treated as a Poisson process, the transition probability in the interval $\delta t$ is $p=1-e^{-\nu\delta t}$, where $\nu$ is the rate for leaving that minimum. The corresponding bidirectional two-level occupation probability is summarized in Appendix~\ref{app:two_level}, and practical transition-probability estimates of this kind are widely used~\cite{chantrell2000calculations,suess2007reliability}.

It is important to note that the generalized SW model has also been implemented and used in several studies as kinetic Monte Carlo (kMC) approaches~\cite{chantrell2000calculations,tan2014magnetic,ruta2015unified,jonasson2019modelling,wolfschwenger2024molecular}. In a kMC approach, the magnetic degrees of freedom are governed by sampling the magnetic energy of the SW model through an MC algorithm. The transition probabilities are calculated in accordance with the generalized SW model. kMC schemes are, however, not limited to the two-state approximation in which case the dipole moment does not necessarily point along the minima of the magnetic energy.

\paragraph{Limits and Performance Considerations}

The validity of the tSW model has been tested against accurate numerical solutions of the Egg model~\cite{mostarac2025thermal}. The precise regime of validity depends on the magnetic field strength and ratio of relaxation times. As a rule of thumb, the tSW model was seen to give quantitatively accurate results for anisotropy parameters $\sigma\gtrsim 5$. If one does away with the two-state approximation, kMC schemes extend the validity of the Stoner-Wohlfarth model towards lower $\sigma$ values, at a substantial computational cost. If one is considering resorting to kMC schemes, we suggest considering the Egg model (see Section~\ref{subsec:egg}) instead.

In the tSW model (and kMC schemes) the dipole moment in an MNP experiences discrete jumps, meaning that the dipole moment orientation is reset at every integration of the particle rotational degrees of freedom. This means that the dipole moment ``trajectory'' is not continuous, and the overall magnetic response of the system is instantaneous. This has important practical implications. One cannot use the standard Green-Kubo relations to, for example, obtain initial dynamic susceptibilities with this method. Furthermore, as a result of the assumptions inherent to the SW model, the real part of the initial dynamic susceptibility will not go to 0 for infinitely high frequency, but to a fixed non-zero value~\cite{mostarac2025thermal}.

Numerically, the accuracy and performance of the tSW approach is mainly related to the minimization strategy used. Depending on the required accuracy for the minimizer, one can trade some precision for performance. The model can be used with multiple integration schemes and it does not need to evaluate the internal magnetic relaxation synchronously with them, which is an important point of flexibility.

\subsubsection{Diffusion-Jump Model}\label{subsec:DJ}

For MNPs with large anisotropy parameters $\sigma$ defined in Eq.~\eqref{anisotropy-parameter}, N\'eel relaxation via thermally activated magnetization reversals becomes a rare event.
The same situation is considered in the tSW model above.
Here, we consider sufficiently large anisotropy energies that deviations of the magnetization direction from the particles' easy axes are small, allowing the magnetization direction to be eliminated as a separate degree of freedom. The resulting simplification leads to the diffusion-jump model~\cite{ilg2019diffusionjump}, with extensions to interacting systems~\cite{pimk_FFBDMC,ilg2023optimality}.
The numerical implementation is described in detail in these references.

One way to think about this is as a simplification of the tSW model when the two local energy minima are constrained to the easy axis direction, with the corresponding barriers entering the jump rate.
Another way is to view the diffusion-jump model as an extension of the rigid-dipole approximation that is supplemented by a Poisson jump process which flips the magnetic moments between $\pm\vec{e}$ with a rate constant proportional to $1/\tauN$.
This formulation allows us to also obtain some analytical results on first-order corrections to the rigid-dipole approximation due to N\'eel relaxation~\cite{ilg2024nonequilibrium,ilg2024stochastic}.

\paragraph{Limits and Performance Considerations}

The validity of the diffusion-jump model has been carefully tested against accurate numerical solutions of the egg model~\cite{ilg2022longest}.
Unsurprisingly, the model is found to be limited to magnetically hard MNPs with rather large anisotropy parameters. The precise regime of validity depends on the magnetic field strength and ratio of relaxation times. As a rule of thumb, the diffusion-jump model was seen to give quantitatively accurate results for anisotropy parameters $\sigma\gtrsim 8$.

\subsubsection{Coupled LLG}\label{subsec:macrospin_llg}\label{subsec:coupled_draft}

The most general continuous description of the coupled magneto-mechanical dynamics of a single-domain MNP combines the mechanical equation of motion for the particle orientation
with the stochastic Landau-Lifshitz-Gilbert (LLG) equation for the magnetization. 
Motivated by the Bloch equations and using methods of irreversible thermodynamics, the classical Landau-Lifshitz equation was originally developed for the magnetization dynamics of a ferromagnetic crystal, but is equally applicable to a ferromagnetic body with uniaxial magnetic anisotropy~\cite{landau_theory_1935}.
Later, Gilbert proposed an equivalent formulation of the Landau-Lifshitz equations starting from a different phenomenological damping term~\cite{Gilbert1955}.

The LLG equation for the magnetization direction $\bmag$ of an individual MNP is
\begin{equation}
\dot{\bmag} = - \frac{\mu_0 \gamma}{1+\alpha^{2}} \bmag \times \vec{H}_{\text{eff}} - \frac{\alpha \mu_0 \gamma}{1+\alpha^{2}} \bmag \times (\bmag \times \vec{H}_{\text{eff}}),
\label{eq:LLG}
\end{equation}
where $\gamma$ is the gyromagnetic ratio, $\alpha$ the Gilbert damping parameter, and $\vec{H}_{\text{eff}}$ the effective magnetic field.
We note in passing that some authors introduce a renormalized gyromagnetic ratio $\gamma/(1+\alpha^2)$. This choice can be justified when starting from Gilbert’s equation \cite{garcia-palacios_langevin-dynamics_1998}.
The first term in Eq.~\eqref{eq:LLG} describes Larmor precession; the second describes damping toward the effective-field direction.
The fundamental timescale resolved by this equation is the microscopic attempt time~(\ref{eq:tau0})~\cite{Wernsdorfer:1997vk,coffey_thermal_2012}.

{In the fully coupled model, the effective field comprises the energy-derived, thermal, and {Barnett} contributions,}
\begin{equation}
{\vec{H}_{\text{eff}} = \vec{H}_{E} + \vec{H}_{\text{therm}} + {\vec{H}_{\text{Bar}}}.}
\label{eq:Heff_combined}
\end{equation}

\noindent
The energy-derived contribution to the effective field for a homogeneously magnetized particle is 
\begin{equation}
{\vec{H}_{E} = -\frac{1}{\mu_0 M_s V_m} \frac{\partial E}{\partial \bmag},}
\end{equation}

\noindent
where $E$ is the total magnetic energy defined in Sec.~\ref{sec::relevant_scales}.  
For the uniaxial macrospin considered here,

\begin{equation}
    \vec{H}_{E}=\vec{H}_{\text{ext}}+\vec{H}_{\text{ani}},
\end{equation} 
where the anisotropy contribution is

\begin{equation}
\vec{H}_{\text{ani}}=\frac{2K}{\mu_0 M_s}(\bmag\cdot\baxis)\,\baxis.
\end{equation}

\noindent
Thermal fluctuations enter through a stochastic contribution to the effective field~\cite{brown1963,kubo_brownian_1970,garcia-palacios_langevin-dynamics_1998},
\begin{equation}
\vec{H}_{\text{therm}} = \vec{\beta} \sqrt{ \frac{2 k_{B} T \alpha}{\gamma \mu_{0}^2 M_{s} V_{m} \Delta t} },
\label{eq:h_therm}
\end{equation}
with $\vec{\beta}$ being a standard-normal random vector, satisfying the fluctuation-dissipation theorem
\begin{equation}
  \langle \vec{H}_i^\text{therm}(t) \vec{H}_j^\text{therm}(t') \rangle = \frac{2 k_{B} T \alpha}{\mu_0^2 M_{s} \gamma V_{m}} \delta_{ij} \delta(t - t').
\end{equation}

\noindent
The Barnett field $\vec{H}_{\text{Bar}}$ arises from mechanical rotation of the particle
and is given by~\cite{barnett_magnetization_1915}
\begin{equation}
{\vec{H}_{\text{Bar}} = -\frac{\alpha}{\gamma \mu_{0}} \bmag \times \bomega,}
\label{eq:barnett}
\end{equation}

The mechanical rotation of the spherical particle is governed by conservation of angular momentum~\cite{keshtgar_magnetomechanical_2017,usadel_dynamics_2015,usadel_dynamics_2017,durhuus_conservation_2024,helbig_self-consistent_2023},
\begin{equation}
J \dot{\bomega} =
\frac{M_{s}V_{m}}{\gamma} \dot{\bmag}
+ \mu_{0}M_{s}V_{m}\, \bmag \times \vec{H}_{\text{ext}}
-\zeta^r \bomega + \bT^\mathrm{therm},
\label{eq:mechanical}
\end{equation}

\noindent
where the first term on the right-hand side is the Einstein-de-Haas torque~\cite{einstein_experimental_1915} coupling spin and lattice angular momentum, the second term is the Zeeman torque, $\zeta^r$ is the rotational friction coefficient, and $\bT^\mathrm{therm}$ a thermal stochastic torque.
The particle orientation, defined by the easy-axis direction $\baxis$, follows the kinematic relation
\begin{equation}
\dot{\baxis} = \bomega \times \baxis.
\label{eq:rotkin}
\end{equation}
Mechanical rotation also feeds back into the LLG through the Barnett field,
which, together with the Einstein-de-Haas torque, ensures consistent transfer of angular momentum between the magnetic and mechanical subsystems.
As shown in Appendix~\ref{app:angmom}, the total angular momentum
\begin{equation}
\vec{L} = J \bomega - \frac{M_s V_m}{\gamma}\bmag
\end{equation}

\noindent
is conserved with respect to all internal magneto-mechanical processes. It changes only due to external torques (Zeeman, viscous friction, and thermal noise).

\paragraph{Limits and Performance Considerations}

The described fully coupled system resolves the fast Larmor precession and its damping, characterized by $\omega_L$ and $\tau_0$, as well as rotational diffusion ($\tauD$), Brownian ($\tauB$), and N\'eel ($\tauN$) relaxation (see Fig.\ \ref{fig:perspectives_overview_doodle}). It is therefore, in principle, applicable across all anisotropy regimes $\sigma$. However, resolving the fast precessional dynamics while reaching the exponentially growing N\'eel relaxation time makes the model computationally extraordinarily expensive at large~$\sigma$.

\textit{Time Integration.} The mechanical equation of motion~\eqref{eq:mechanical} retains rotational inertia and is second order in the particle orientation, 
requiring an integrator suitable for Langevin dynamics.
In ESPResSo, this is handled by the Velocity Verlet algorithm operating on quaternion-based particle orientations with a fixed time step~\cite{rapaport_art_2004}.
The LLG equation~\eqref{eq:LLG}, being first order, is integrated in tandem using an adapted Heun's algorithm \cite{butcher2000numerical} that achieves comparable accuracy to the Velocity Verlet scheme.
Within each time step, the Verlet half-step first advances velocities and positions, after which the Heun predictor-corrector updates the magnetization direction $\bmag$, and finally, the Verlet step is completed with the new forces and torques.
A known limitation of this interleaved scheme is that, for full consistency, a predictor for the mechanical state would be needed within the intermediate Heun steps. This is omitted because the characteristic times of fast magnetic and slow mechanical rotation are typically well separated.

\subsubsection{Egg Model}\label{subsec:egg}
For many practical situations, the Larmor precession and
magneto-mechanical inertial effects occur on timescales much faster than those of interest.
An approach for simulating coupled MNP dynamics that makes use of this separation is the so-called ``egg model'' (EM).
It was originally introduced by Shliomis and Stepanov to describe non-equilibrium magnetic response of dilute ferrofluids~\cite{shliomis_theory_1994}.
Recently, the model has gained increasing
popularity as a robust tool to model superparamagnetic
dynamics~\cite{taukulis2012coupled,poperechny_combined_2020,kroger_combined_2022,poperechny2023multipeak,ilg2022longest,PYANZINA2025,Khelfallah2026}.

Within the egg model, the mechanical rotation of a spherical single-domain MNP immersed in a viscous matrix is governed by the standard kinematic relationship Eq.~\eqref{eq:rotkin} with angular velocity given by

\begin{equation}
\bomega = \frac{\mu_0 M_s V_m}{\zeta^r} \bmag \times \vec{H}_\text{ext} + \bomega_\text{therm}, \label{eq:egg_omega}
\end{equation}

\noindent where $\bomega_\text{therm} = \bT^\mathrm{therm}/\zeta^r$.

Rotation of the magnetic moment in the body-fixed frame (bff) of the particle is modeled as

\begin{gather}
\frac{d\bmag}{dt}\bigg|_\text{bff} = \left(\bomega^{L} + \bomega^{m} \right)\times \bmag, \label{eq:egg_kin} \\
\bomega^{L} = \frac{\mu_0 \gamma}{1 + \alpha^2} \vec{H}_\text{E}, \\
\bomega^{m} = \frac{\mu_0 \mu}{\zeta^m} \bmag \times \vec{H}_\text{E} + \bomega_\text{therm}^m , \label{eq:egg_omega_e}
\end{gather}

\noindent where $\bomega^{L}$ is the Larmor precession frequency,
$\bomega^{m}$ is the dissipative component of the magnetic angular velocity,
$\zeta^m$ is the ``magnetic friction'' coefficient,

\begin{equation} \label{eq:zeta_m}
\zeta^m = \frac{\mu (1 + \alpha^2)}{\alpha \gamma} = 2 \tauD k_B T .
\end{equation}

\noindent
Fluctuations $\bomega_\text{therm}^m$ are modeled as white noise with zero mean and autocorrelation function

\begin{gather} \label{eq:egg_therm_omegas}
\langle \omega_{\text{therm},i}^{m}(t) \omega_{\text{therm},j}^ m(t') \rangle = \frac{1}{\tauD} \delta_{ij} \delta(t - t').
\end{gather}

\noindent
{The transition from the body-fixed frame to the space-fixed frame is performed as}

\begin{equation} \label{eq:frame_transition}
\dot{\bmag} = \frac{d\bmag}{dt}\bigg|_\text{bff} + \bomega \times \bmag.
\end{equation}

In practice, the Larmor precession is often neglected within the EM,
\textit{i.e.}, one sets ${\omega^L = 0}$~\cite{taukulis2012coupled,kroger_combined_2022}.
In this case,
the form of Eqs.~\eqref{eq:egg_kin} and \eqref{eq:egg_omega_e} closely mirrors that of Eqs.~\eqref{eq:rotkin} and \eqref{eq:egg_omega}.
Thus, the magnetization dynamics within the particle body can be metaphorically interpreted as a rotation of a magnetic sphere (`` egg yolk'') in an effective viscous medium of viscosity $\eta^m = \zeta^m/6 V_m$ (``egg white''). Hence, the model name.

\paragraph{Comparison to the Coupled LLG}

Let us emphasize similarities and differences between the
EM and the coupled LLG model
presented in Sec.~\ref{subsec:macrospin_llg}:

\begin{itemize}[leftmargin=*]
\item Equation~\eqref{eq:egg_omega} for the MNP mechanical rotation directly follows from the angular momentum conservation law Eq.\ \eqref{eq:mechanical} upon assuming the limit $\dot{L} \rightarrow 0 $.
Thus, both the rotational inertia and the Einstein-de-Haas effect are neglected within the EM;

\item The effect of noise on internal magnetization dynamics
is incorporated differently within the two models.
The LLG equation~\eqref{eq:LLG} uses the thermal field $\vec{H}_{\text{therm}}$,
which enters both precession and damping terms,
while the EM adds noise via $\bomega^m_{\text{therm}}$,
separately from the damping term.
In both cases, the noise satisfies the fluctuation-dissipation theorem and should result in identical ensemble-averaged properties~\cite{garcia-palacios_langevin-dynamics_1998};

\item The effect of mechanical rotation on $\bmag$
is modeled differently as well.
To show it, let us consider the athermal limit $T = 0$ and rewrite
the LLG~\eqref{eq:LLG} as

\begin{gather}
  \dot{\bmag} = \frac{d\bmag}{dt}\bigg|_\text{bff} + \frac{\alpha}{1 + \alpha^2} \bmag \times \left[ \bmag \times \bomega - \alpha \bomega \right]. \label{eq:llg_bff_form}
\end{gather}

\noindent
Here, the first term on the r.h.s. is the same as in the EM and is given by Eq.~\eqref{eq:egg_kin}.
It is independent of $\bomega$.
The second term follows from the Barnett field definition [Eq.~\eqref{eq:barnett}].
Comparing Eq.~\eqref{eq:llg_bff_form} to \eqref{eq:frame_transition},
one can see that $\omega$-dependent terms in the two models only coincide in the high damping limit
$\alpha \rightarrow \infty$.
At $\alpha = 0$,
the explicit $\omega$-dependent coupling in the LLG vanishes.
But within the EM, this coupling exists independently of the Gilbert damping.
\end{itemize}

\paragraph{Limits and Performance Considerations}

The coupled LLG model, Eqs.~\eqref{eq:LLG}--\eqref{eq:barnett}, is needed when precessional dynamics, inertial effects, or magneto-mechanical coupling at short timescales are relevant --
for example, in ferromagnetic resonance or when the Einstein-de-Haas and Barnett effects play a role, as in the case of ultra-low viscosity~\cite{durhuus_conservation_2024,keshtgar_magnetomechanical_2017}.
The simpler egg model, Eqs.~\eqref{eq:egg_omega}--\eqref{eq:frame_transition}, is the preferred choice when these fast processes can be safely averaged over, offering a significant reduction in computational cost while still capturing the essential physics of combined Brown--N\'eel relaxation.
For both models, the exponential growth of $\tauN$ at large $\sigma$ eventually necessitates the coarse-grained approaches discussed above.
If the homogeneous single-domain approximation fails, the models above are no longer sufficient and one must switch to a spatially resolved micromagnetic description.

An ESPResSo implementation of the egg model was proposed in Ref.~\cite{PYANZINA2025}.
There, the equation for the magnetization dynamics within the particle frame, Eq.~(\ref{eq:egg_kin}),
was integrated using the package's standard Brownian Dynamics integrator.
The magneto-mechanical coupling described by Eq.~(\ref{eq:frame_transition}) was realized using
the ``virtual sites'' feature of ESPResSo and the raspberry model (see Sec.~\ref{subsec::hydrodynamics}).
This approach allows one to simulate in a straightforward manner not only single-domain MNPs, but also multicore particles with an arbitrary internal arrangement of magnetic grains.

\subsection{Multidomain Particles}\label{subsec:mmodel_multi}

Magnetic particles that form domains, domain walls, vortices, antiparallel layers and other inhomogeneous magnetization states can be well described within the framework of micromagnetism~\cite{abert_micromagnetics_2019}.
Micromagnetism treats the magnetization as a continuous vector field of fixed magnitude, whose equilibrium states are obtained by minimizing the total magnetic free energy and whose time evolution is governed by the Landau--Lifshitz--Gilbert equation. This is the regime in which the homogeneous single-domain models above cease to be sufficient: a weak or vanishing particle moment can now arise from spatially nonuniform magnetization within the particle rather than from the temporal averaging of a single macrospin.

Superparamagnetic iron-oxide nanoparticles (SPIONs) are commonly employed to mitigate magnetic agglomeration arising from interparticle interactions. Owing to thermal fluctuations, the magnetic moments of SPIONs rapidly reverse, resulting in a vanishing time-averaged magnetization in the absence of an external field. As a consequence, long-lived dipolar interactions are suppressed, reducing the tendency for particles to agglomerate. However, superparamagnetism is not an absolute material property but depends on the experimental timescale. A particle behaves as superparamagnetic if its Néel relaxation time $\tauN$ is shorter than the characteristic timescales associated with particle motion, namely translational diffusion $\tau_{\mathrm{diff}}$ and rotational Brownian motion $\tau_B$~\cite{serantes2021nanoparticle}. When $\tauN < \tau_{\mathrm{diff}}$ and $\tauN <\tauB$, the magnetic moment fluctuates sufficiently fast to prevent stable magnetic coupling between particles, thereby minimizing agglomeration even below the blocking temperature.

In contrast to superparamagnetism, where rapid thermal fluctuations of the magnetic moment result in a vanishing magnetization when averaged over time, zero net magnetization can also be achieved through a spatial averaging mechanism, in which the magnetization cancels within the particle itself.
Examples are flux-closure states, such as magnetic vortices, in which the magnetization curls on the length scale of the particle. In these configurations the stray field is strongly suppressed, except for a small dipolar contribution originating from the vortex core. Building on this concept, Kim \textit{et al.} demonstrated the use of biofunctionalized magnetic microdiscs with a spin-vortex ground state for targeted cancer-cell destruction~\cite{kim2010biofunctionalized}.
Another approach to reducing interaction fields is the use of synthetic antiferromagnetic particles, in which thin magnetic layers are coupled either via dipolar interactions or via interlayer exchange coupling, resulting in an antiparallel alignment of the magnetization~\cite{welbourne2021high,D5BM00739A}.

\subsubsection{Micromagnetics with Mechanical Rotation}

For particles above the single-domain limit, the macrospin approximation is no longer appropriate, and micromagnetic modeling must be used to resolve the magnetic domains. Zehner~\cite{zehner_integrating_2024} combined the molecular dynamics framework {ESPResSo}~\cite{weik19a,grad_2026_20450775} with the finite-difference micromagnetic Python library \textit{magnum.np}~\cite{bruckner_magnumnp_2023} to simulate the behavior of a multidomain magnetic particle in a rotating magnetic field. The approach used a raspberry representation, introduced in Section \ref{subsec::hydrodynamics}.

In this generalized model, the LLG equation (\ref{eq:LLG}) is solved for the magnetization unit vector $\bmag(\vec{x})$ in each discretization cell of the regular finite-difference mesh, with the effective field $\vec{H}_{\text{eff}}(\vec{x})$ also including exchange and dipolar interactions. The Zeeman, anisotropy, and Barnett fields are unaltered and evaluated locally. Regarding thermal fluctuations, an appropriate scaling must be applied in order to yield the correct thermal statistics. A common choice is a random vector field, proposed in~\cite{martinez_micromagnetic_2007}, and given by
\begin{equation}
\vec{H}_{\text{therm}} = \vec{\beta} \sqrt{ \frac{2 k_{B} T \alpha}{\gamma \mu_{0}^2 M_{s} \Delta x^3 \Delta t} }\,\,,
\end{equation}
where $\Delta x$ is the linear cell size used in the micromagnetic simulation, as opposed to the particle volume appearing in the single-domain thermal field of Eq.~\eqref{eq:h_therm}.

In the mechanical equation of motion~(\ref{eq:mechanical}), the macrospin $\bmag$ is replaced by the volume average of the normalized magnetization field:
\begin{equation}
\overline{\bmag} = \frac{1}{V_m}\int_{V_m}\bmag(\vec{x})\,\mathrm{d}V.
\end{equation}
This substitution is justified by the linear dependence of the external-field torque on the magnetization, which allows its effect to be expressed through the volume-averaged magnetization. An analogous argument holds for the Einstein--de Haas term, since the total angular momentum transfer is the sum of the contributions from the individual magnetic moments.

Finally, the orientation of the uniaxial anisotropy axis (Eq.~\eqref{eq:rotkin}) is no longer a sufficient coordinate to fully describe the particle orientation. If the magnetization configuration lacks rotational symmetry about the anisotropy axis, rotations around this axis lead to physically distinct states and must therefore be included as additional orientational degrees of freedom. This can be achieved by describing the particle's orientation with a rotation matrix $\boldsymbol{R}$ or a quaternion coordinate $\boldsymbol{q}$. In the first case, the evolution of $\boldsymbol{R}$ is given by the relation
\begin{equation}
\dot{\boldsymbol{R}}= [\bomega\times]\boldsymbol{R}\,,
\end{equation}
where $[\bomega\times]$ is derived from the angular velocity of the particle and denotes the skew-symmetric so-called cross product matrix
\begin{equation}
[\bomega\times]=
\begin{pmatrix}
0&-\omega_3&\omega_2\\
\omega_3&0&-\omega_1\\
-\omega_2&\omega_1&0
\end{pmatrix}.
\end{equation}
While the description using a rotation matrix is more intuitive and convenient for analytical considerations, the quaternion formalism is typically favored in numerical simulations due to its improved numerical stability and is used in the present integration scheme.

In the present combined micromagnetic simulations, no additional particles are taken into account. For the intended single-particle study, the internally resolved micromagnetic object is, therefore, represented as one rigid particle within the MD framework, without replacing its magnetization texture with a macrospin. The absence of additional particles restricts the demonstrated setup to isolated-particle or strongly dilute systems.
Compared with the single-domain models discussed above, when dealing with an ensemble of interacting multidomain particles, the computational cost rises significantly, since the magnetization dynamics have to be computed in every discretization cell instead of only for a single particle moment.

\section{Dipolar Coupling in Colloidal Simulations}\label{sec::dipdip}
The model hierarchy above sets the internal magnetic description. Dipolar coupling, introduced in Section~\ref{sec::relevant_scales} via Eq.~\eqref{eq:dipole-dipole}, is an additional many-body ingredient that links the integration scheme to the magnetodynamic models in Section~\ref{sec::mmodels}. Like the hydrodynamic couplings of Section~\ref{subsec::hydrodynamics}, it is a long-range interaction that shapes collective behavior across model classes without by itself determining the internal magnetization dynamics. Including dipole-dipole interactions in many-body colloidal simulations is computationally demanding because of their long range: naïve direct summation scales as $O(N^2)$ and is prohibitive for large $N$. To overcome this, methods such as Ewald summation, Particle-Particle Particle-Mesh (P3M), and Fast Multipole Methods (FMM) reduce the cost to $O(N^{3/2})$, $O(N \log N)$ or even $O(N)$, enabling simulations of much larger dipolar systems \cite{2008-cerda-jcp}.

For the fixed-point-dipole, LLG and Egg models, Eq.~\eqref{eq:dipole-dipole} enters directly as the pair interaction potential in equations of motion (Eqs.~\eqref{eq:maisF} and ~\eqref{eq:IdwisT}), and explicit long-ranged dipolar solvers under open, periodic, or mixed boundary conditions become part of the practical simulation methodology. This means that for these models, the dipole-dipole interactions are explicitly resolved outside the respective magnetodynamic models. In the LLG model, however, the local dipolar field must additionally be available to evaluate the precessional dynamics. In general, the total local dipolar field is already evaluated by dipole-dipole interaction solvers as part of the standard torque calculation in Eq.~\eqref{eq:IdwisT}. In the effective-field method, by contrast, interparticle interactions are not resolved pairwise but can be incorporated within the framework of dynamic mean-field theories~\cite{ilg_magnetoviscosity_2005,Fang2022}.

The Stoner-Wohlfarth model is designed for non-interacting systems. However, it is possible to implicitly include dipole-dipole interactions by redefining the input magnetic field in the model. The idea is that each MNP is under the combined influence of applied magnetic fields and the total dipolar fields. Provided there is a way to estimate the total dipolar field at the particle position, it is sufficient to use the total field acting on the particle rather than just the applied field when using the model. For homogeneous, semi-dilute systems, this can be done analytically. Using P3M methods~\cite{2008-cerda-jcp}, in combination with hardware accelerators, the tSW model becomes a high-performance, scalable method to incorporate N\'eel relaxation also for thermal, interacting systems.

Analogous remarks hold for the diffusion-jump model, in which N\'eel relaxation is realized even more efficiently, but which is restricted to magnetically hard MNPs. Recently, initial simulation investigations of interacting dipolar systems that incorporate both Brownian and N\'eel relaxation through the diffusion-jump approach have been carried out, including evaluations of appropriately modified dynamical mean-field theories \cite{pimk_FFBDMC,ilg2023optimality}.

\begin{figure*}[t]
  \centering
\subfigure[]{\label{fig:micro_rasp_mesh}\includegraphics[width=0.24\linewidth]{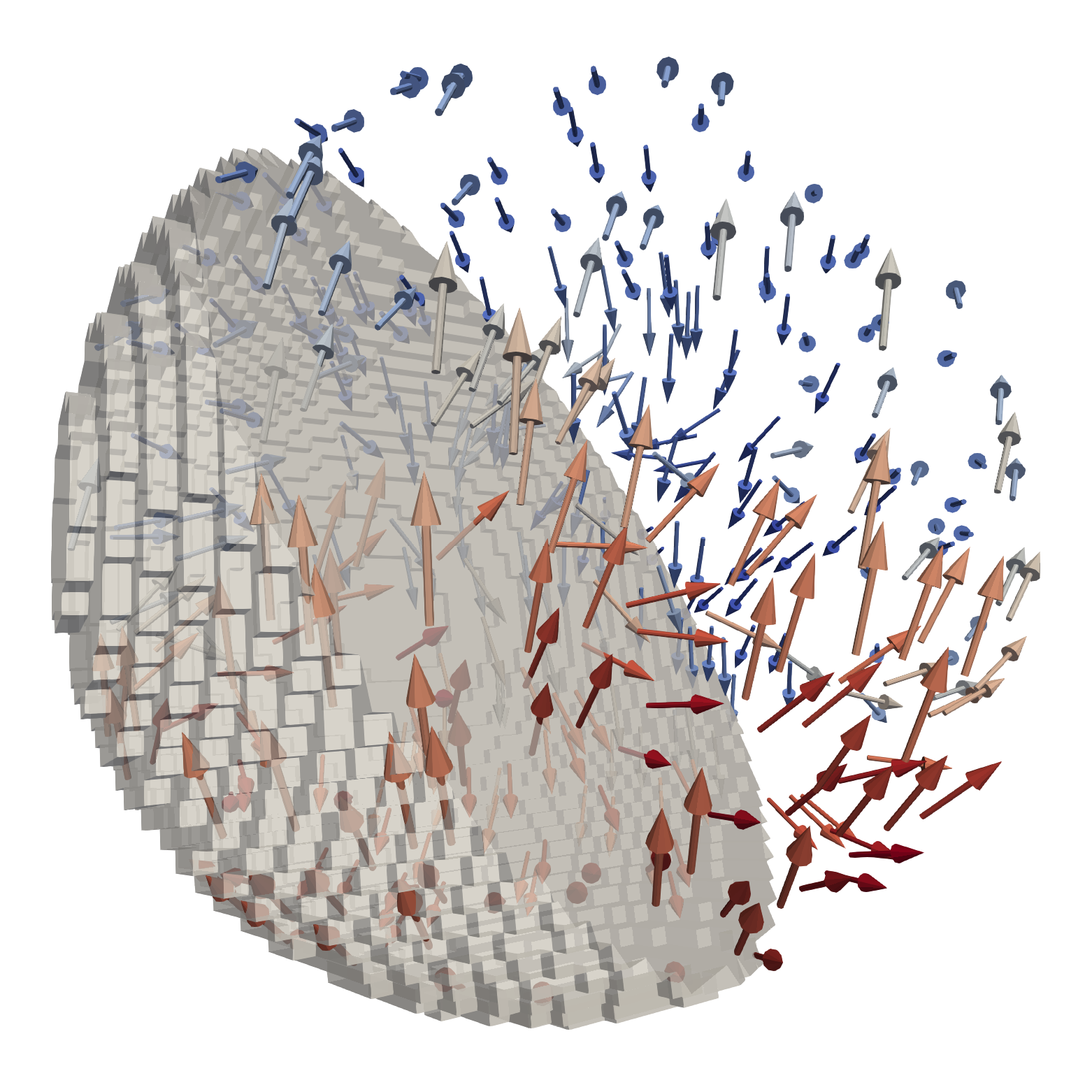}}
\subfigure[]{\label{fig:dist_dipl_rasp}\includegraphics[width=0.24\linewidth]{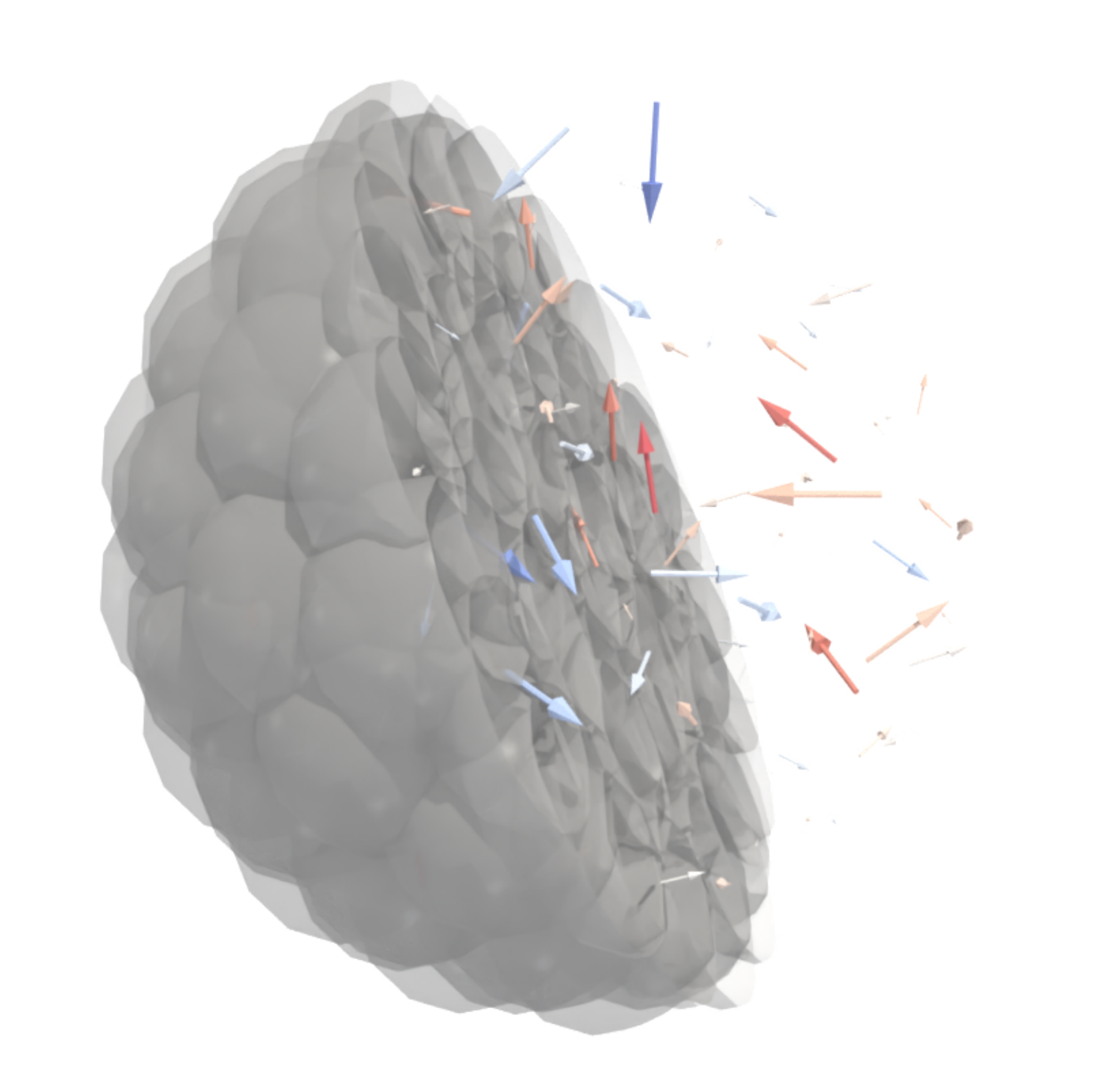}}
\subfigure[]{\label{fig:point_dip_rasp}\includegraphics[width=0.49\linewidth]{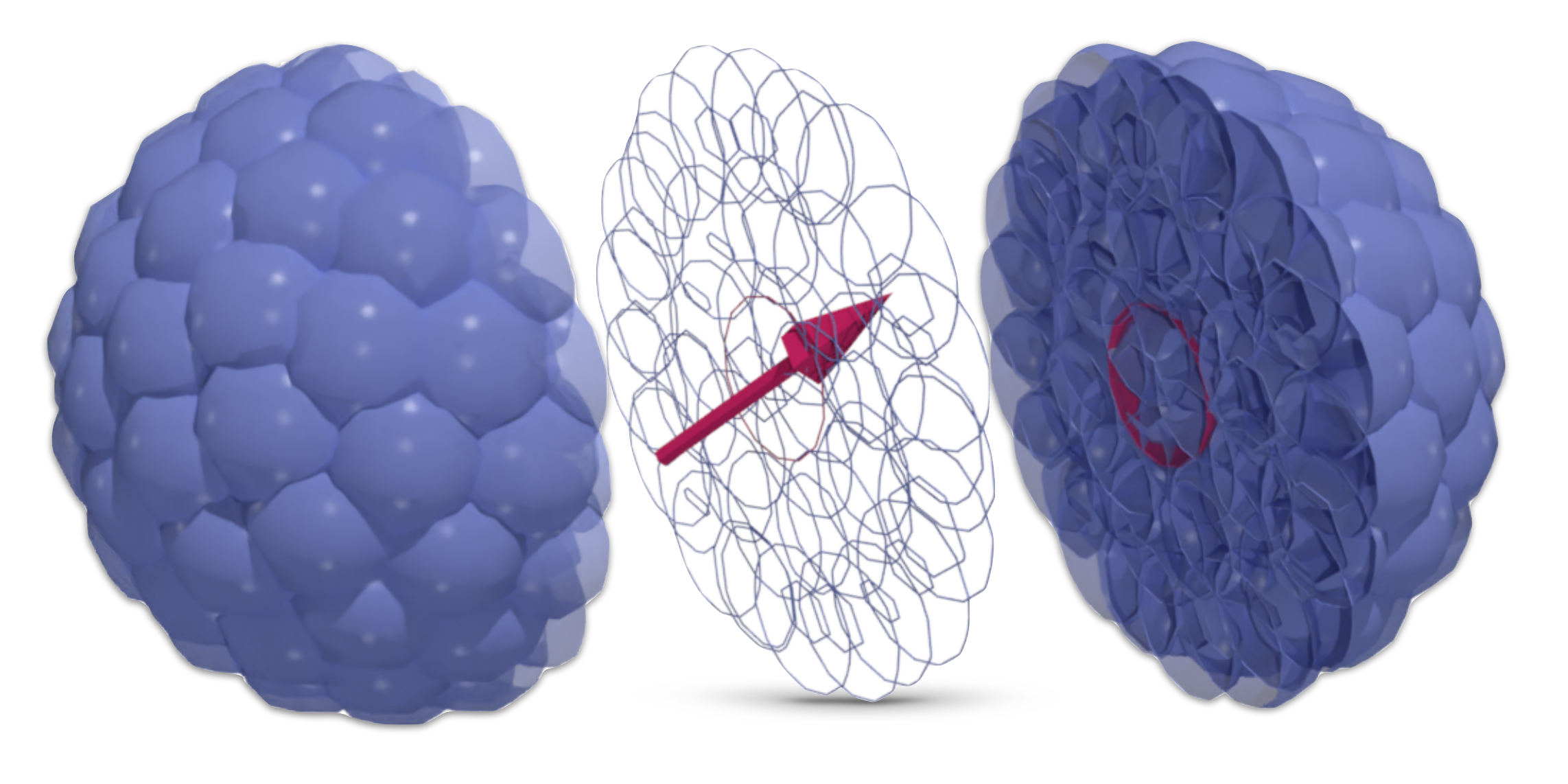}}

  \caption{Rendered simulation representations of MNPs with their internal dipole structures annotated. \ref{fig:micro_rasp_mesh} shows the micromagnetic mesh of an MNP with its internal cell moments. \ref{fig:dist_dipl_rasp} shows a raspberry representation with distributed dipole moments. Their positions within the raspberry are chosen here to reflect the frictional coupling scheme used with the lattice-Boltzmann method; this arrangement is illustrative rather than a general requirement. \ref{fig:point_dip_rasp} shows a raspberry MNP whose magnetic texture is mapped onto a point dipole at its center, forming an effective point-dipole approximation.}
  \label{fig:micromag_rasp}
\end{figure*}

When modeling an ensemble of interacting multidomain particles using full micromagnetic simulations, two coupling schemes can be employed, as illustrated in Fig.~\ref{fig:micromag_rasp}. In the first approach, the micromagnetic mesh shown in Fig.~\ref{fig:micro_rasp_mesh} is directly mapped onto the distributed dipoles of the raspberry model (Fig.~\ref{fig:dist_dipl_rasp}). This enables the explicit calculation of all pairwise dipolar interactions between all mesh elements of all particles in the ensemble. The resulting forces and torques are then mapped back onto the particles' centers of mass, while the local magnetic fields are recalculated from the updated dipole configuration. Alternatively, for the calculation of interparticle dipolar forces and torques, each micromagnetically resolved multidomain particle can be coupled to the colloidal solver through the volume-averaged effective moment $\vec{\mu}_{\mathrm{eff}}=M_sV_m\overline{\bmag}$, as illustrated in Fig.~\ref{fig:point_dip_rasp}. This effective point-dipole coupling reduces the computational cost without replacing the internal micromagnetic texture, which continues to evolve on the cell mesh. It neglects the spatial distribution of the magnetization only in the interparticle interaction calculation, making it less accurate, particularly when short-range interactions or densely packed multidomain particles are important. The choice between the two approaches is therefore a trade-off between accuracy and computational efficiency. While the effective point-dipole coupling scales well to large particle ensembles, the fully resolved dipolar approach rapidly becomes computationally prohibitive as the number of particles increases.

\section{Conclusion}

In this Perspective, we have reviewed the principal modeling approaches for magnetic nanoparticle dynamics across different physical regimes, ranging from coarse-grained macrospin descriptions to fully resolved micromagnetic simulations. We have highlighted how the choice of an appropriate model is governed by the relevant energy and timescales. Rather than advocating a single universal framework, we have emphasized that the optimal level of description depends on the specific material properties, particle size, interaction strength, and physical processes relevant to the application. Particular attention was given to the coupling between magnetization dynamics, translational and rotational particle motion, hydrodynamic interactions, and long-range dipolar interactions, all of which are essential for realistic simulations of magnetic nanoparticle suspensions. The broad range of available methods reflects the multiscale nature of magnetic nanoparticle systems. While simplified models enable simulations of large ensembles over experimentally relevant timescales, more detailed descriptions are required whenever internal magnetization dynamics or multidomain effects become important. Consequently, selecting an appropriate model requires balancing physical fidelity against computational efficiency. Table~\ref{tab:model_selection_summary} summarizes the regimes of applicability discussed throughout this article and provides a practical guide to model selection. Although considerable progress has been made, important challenges remain before predictive multiscale simulations of realistic magnetic nanoparticle systems become routine. These challenges and promising future research directions are discussed in the following Outlook.

\begin{table*}[htbp]
\scriptsize
\centering
\caption{Practical summary of the magnetic models discussed in
Section~\ref{sec::mmodels}.}
\label{tab:model_selection_summary}

\renewcommand{\arraystretch}{1.15}
\setlength{\tabcolsep}{4pt}

\begin{tabularx}{\textwidth}{
    >{\RaggedRight\arraybackslash}p{0.13\textwidth}
    >{\hsize=1.25\hsize}Y
    >{\hsize=1.15\hsize}Y
    >{\hsize=0.75\hsize}Y
    >{\hsize=0.75\hsize}Y
}
\toprule
Model
&
Typical use case / regime
&
Characteristics
&
Representative references
&
Implementation available in:
\\
\midrule

Fixed-point dipole
&
Single-domain MNPs with frozen internal magnetization; rigid-dipole approximation
&
Resolves $\tau_{\mathrm B}$; no internal magnetization dynamics; no computational overhead
&
\cite{ilg_magnetoviscosity_2005,Ilg_lnp,Sreekumari2013,Satoh99,cerda_aggregate_2008}
&
ESPResSo release, LAMMPS, InnMNP
\\
\addlinespace[0.5em]

Effective field method
&
Ensemble-level description of dilute, non-interacting single-domain MNP suspensions
&
Rigid-particle Brownian orientational dynamics; ensemble-averaged magnetization
&
\cite{martsenyuk1974kinetics,raikher1994effective,Shliomis2001,Rinaldi2010}
&
---
\\
\addlinespace[0.5em]

Thermal Stoner--Wohlfarth
&
Single-domain MNPs with $\sigma \gtrsim 5$
&
Resolves $\tau_{\mathrm B}$ and $\tau_{\mathrm N}$; N\'eel relaxation through thermally activated switching between minima; low computational overhead
&
\cite{stoner_mechanism_1948,chuev_nanomagnetism_2007,mostarac2025thermal,wolfschwenger2024molecular}
&
ESPResSo release, InnMNP
\\
\addlinespace[0.5em]

Diffusion-jump model
&
Single-domain MNPs with $\sigma \gtrsim 8$, when the dipole orientation does not deviate significantly from the easy axis
&
Resolves $\tau_{\mathrm B}$ and $\tau_{\mathrm N}$; N\'eel relaxation represented as rare jump events; low computational overhead
&
\cite{ilg2019diffusionjump,pimk_FFBDMC,ilg2023optimality,ilg2022longest}
&
---
\\
\addlinespace[0.5em]

Stochastic LLG
&
Most general continuous model of magnetization dynamics for single-domain MNPs
&
Resolves $\tau_0$, $\tau_{\mathrm D}$, $\tau_{\mathrm B}$, and $\tau_{\mathrm N}$; full coupled dynamics; high computational overhead
&
\cite{landau_theory_1935,Gilbert1955,brown1963,garcia-palacios_langevin-dynamics_1998,usadel_dynamics_2015,durhuus_conservation_2024}
&
ESPResSo developer, InnMNP
\\
\addlinespace[0.5em]

Egg model
&
Continuous model of magnetization dynamics for single-domain MNPs without fast precessional and inertial effects
&
Resolves $\tau_{\mathrm D}$, $\tau_{\mathrm B}$, and $\tau_{\mathrm N}$; full coupled dynamics; moderate computational overhead
&
\cite{shliomis_theory_1994,kroger_combined_2022,PYANZINA2025}
&
ESPResSo developer
\\
\addlinespace[0.5em]

Micromagnetics with mechanical rotation
&
MNPs above the single-domain limit $d>d_{\mathrm{sd}}$, or whenever spatially nonuniform magnetization is important
&
Resolves magnetic relaxation mechanisms; spatial, cell-resolved magnetization; extreme computational overhead
&
\cite{zehner_integrating_2024,bruckner_magnumnp_2023,weik19a,grad_2026_20450775}
&
ESPResSo developer + magnum.np
\\

\bottomrule
\end{tabularx}
\end{table*}

\section{Outlook}

Several important challenges remain unresolved in the modeling of magnetic nanoparticles. While micromagnetic approaches are increasingly employed to describe multidomain particles and complex magnetization states, they are almost exclusively restricted to single-particle or immobilized-particle ensemble studies. The explicit simulation of interacting particles in liquid carriers at the micromagnetic level remains largely unexplored due to the immense computational cost associated with resolving both internal magnetic structures and long-range particle interactions. Bridging this gap, using, for example, a raspberry representation of multidomain magnetic particles, would provide valuable insight into collective phenomena that cannot be captured within macrospin approximations. Another key advantage of the raspberry model is the ability to represent particles of arbitrary shape. Combining this geometric flexibility with micromagnetism opens the prospect of simulating nonhomogeneously magnetized objects, in which both the shape and the internal magnetization distribution of a body are resolved. This would allow the model to capture complex magnetic domains and their coupling to particle geometry, extending the framework beyond the macrospin approximation toward a more faithful representation of real magnetic bodies. Another limitation of current modeling approaches is the treatment of the surrounding carrier medium. Most studies focus on homogeneous carriers and externally applied homogeneous or weakly varying magnetic fields. However, many practical applications involve heterogeneous environments, including biological tissues, porous media, gels, or complex fluids, where local transport properties and magnetic fields can vary substantially in space and time. Extending current models to account for such inhomogeneities represents an important step toward predictive simulations under realistic operating conditions.

The broad range of available magnetic models—from fixed-point dipoles and effective-field methods to stochastic LLG and micromagnetic descriptions—highlights the need for a unified simulation framework.  The ongoing development of ESPResSo, together with its coupling to \textit{magnum.np}, is intended to provide such a framework as an open resource for the community~\cite{weik19a,grad_2026_20450775,zehner_integrating_2024,bruckner_magnumnp_2023}. Most of the models discussed in this Perspective are already implemented in ESPResSo, while \textit{magnum.np} extends the available descriptions to spatially resolved micromagnetics. This combination places the magnetic models within an established particle-based simulation environment that also provides translational and rotational particle dynamics, hydrodynamic interactions, long-range dipolar solvers, and parallel simulation capabilities. A common modular implementation facilitates direct comparisons between models and provides a basis for multiscale coupling between different levels of resolution. This Perspective therefore not only compares the physical regimes of the available models but also states our intent to support their ongoing integration within this unified framework. The coexistence of processes spanning many orders of magnitude in time remains one of the central computational challenges in the field. Fast magnetization dynamics often occur on picosecond timescales, whereas aggregation, phase transitions, transport, and application-relevant processes may evolve over seconds or longer. Efficient multiple-timestep algorithms and adaptive multiscale integration strategies therefore represent a particularly promising direction for future developments. Taken together, the diversity and ongoing development of these models are crucial for reliable predictions of thermomagnetic behavior, transport phenomena, imaging contrast mechanisms, and aggregation dynamics, as well as for identifying the limitations of current theoretical descriptions. Ultimately, integrating these models within a common simulation framework will help connect fundamental model development with the application-driven design of magnetic particles having tailored anisotropy and shape-dependent magnetic responses.

\begin{acknowledgments}
DM was funded by the Austrian Science Fund (FWF) 10.55776/J4915. For open access purposes, the author has applied a CC BY public copyright license to any author accepted manuscript version arising from this submission. SSK was partially supported by the European Union's Horizon Europe research and innovation programme under Marie Skłodowska-Curie Actions Doctoral Network MAESTRI (grant agreement no. 101119614) and the Austrian Science Fund (FWF) [10.55776/PAT4120124, available via https://www.fwf.ac.at/en/discover/research-radar]. SH acknowledges financial support by the Vienna Doctoral School in Physics (VDSP). AAK acknowledges financial support by the Austrian Science Fund (FWF) [10.55776/PAT4307624].
\end{acknowledgments}

\section*{Data Availability}

This is a Perspective article, and no experimental data were analyzed. Numerical curves can be reproduced directly from the equations and parameters reported in the manuscript; no additional datasets were created.

\appendix

\section{Angular Momentum Conservation}\label{app:angmom}

We verify that the full coupled model of Sec.~\ref{subsec:coupled_draft} is consistent with conservation of total angular momentum with respect to internal magneto-mechanical processes. The total angular momentum of the system consists of a mechanical and a spin contribution,
\begin{equation}
\vec{L} = \underbrace{J\bomega}_{\text{mechanical}} - \underbrace{\frac{M_s V_m}{\gamma}\bmag}_{\text{spin}}.
\label{eq:L_total}
\end{equation}
{The spin angular momentum is thus identified as $\vec{S}=-(M_sV_m/\gamma)\bmag$. Because the mechanical equation was constructed from this angular-momentum balance, the calculation below is a consistency check of the coupled equations rather than an independent derivation of the mechanical equation.}
Taking the time derivative,
\begin{equation}
\dot{\vec{L}} = J\dot{\bomega} - \frac{M_s V_m}{\gamma}\dot{\bmag}.
\label{eq:Ldot}
\end{equation}
The mechanical equation of motion~(\ref{eq:mechanical}) contains the Einstein–de-Haas term $\frac{M_s V_m}{\gamma}\dot{\bmag}$, which by construction equals the negative rate of change of the spin angular momentum. Substituting Eq.~(\ref{eq:mechanical}) into Eq.~(\ref{eq:Ldot}), this term cancels the spin contribution identically, leaving
\begin{equation}
\dot{\vec L} = \mu_0 M_s V_m\bmag\times\vec H_{\text{ext}} - \zeta^r\bomega + \vec\tau_{\text{therm}}.
\end{equation}
The right-hand side contains only \emph{external} torques: the Zeeman torque from the applied field, viscous friction with the surrounding fluid, and the thermal stochastic torque.
Crucially, the internal magnetization dynamics---governed by the LLG equation~(\ref{eq:LLG}) including the {Barnett field}---do not appear.
This cancellation holds regardless of the specific form of $\dot{\bmag}$. It is a structural consequence of the Einstein-de-Haas coupling in the mechanical equation.

In the absence of external fields, viscous friction, and thermal noise ($\vec{H}_{\text{ext}} = 0$, $\zeta^r = 0$, $\vec{\tau}_{\text{therm}} = 0$), total angular momentum is strictly conserved: $\dot{\vec{L}} = 0$.

\textbf{{Role of the Barnett field.}}
{While the Einstein-de-Haas term alone guarantees angular momentum conservation, the Barnett field $\vec{H}_{\text{Bar}} = -\frac{\alpha}{\gamma\mu_0}\bmag \times \bomega$ in $\vec{H}_{\text{eff}}$ plays a complementary role: it ensures that mechanical rotation is accounted for in the LLG equation, so that the magnetization dynamics are physically consistent in the rotating frame. Without it, angular momentum would still be formally conserved (since the cancellation of the spin rate against the Einstein–de-Haas term is by construction and does not depend on the form of $\vec{H}_{\text{eff}}$), but the partitioning of angular momentum between spin and lattice degrees of freedom would be incorrect.}

\section{Two-Level Switching Probability}\label{app:two_level}

We assume that each particle can be described by a two-level system. One level corresponds to the state with magnetization up, while the other level corresponds to the state with magnetization down. The occupation probabilities of the two levels, $P_1$ and $P_2$, satisfy the normalization condition
\begin{equation}
P_1 + P_2 = 1,
\end{equation}
and the master equation
\begin{equation}
\frac{dP_1}{dt} = -w_{12} P_1 + w_{21}(1 - P_1).
\label{eq:master_two_level}
\end{equation}

\noindent Here, $w_{12}$ is the switching rate from state 1 to state 2, and $w_{21}$ is the switching rate from state 2 to state 1. We solve this equation assuming that at $t = 0$ the system is in state~1,
\begin{equation}
P_1(t=0) = 1.
\end{equation}

The solution is given by
\begin{equation}
P_1(t) =
\frac{w_{12}}{w_{12} + w_{21}} \,
\mathrm{e}^{-t (w_{12} + w_{21})}
+ \frac{w_{21}}{w_{12} + w_{21}}.
\label{eq:P1_solution}
\end{equation}

The probability of occupying state 2 after a time interval $\Delta t$ is
\begin{equation}
p = 1 - P_1(\Delta t)
= \frac{w_{12}}{w_{12} + w_{21}}
\left[
 1 - \mathrm{e}^{-(w_{12} + w_{21})\,\Delta t}
\right].
\label{eq:switching_prob}
\end{equation}

If $w_{12} \gg w_{21}$, it follows that
\begin{equation}
p \approx 1- \mathrm{e}^{-\Delta t \, w_{12}}.
\end{equation}
Identifying $\nu=w_{12}$ recovers the one-rate expression used in the main text when reverse switching during the interval is negligible. More generally, for a sufficiently short time step, both the exact bidirectional expression and the one-rate formula give $p=w_{12}\Delta t+\mathcal{O}(\Delta t^2)$.

\bibliography{bibliography}

@article{BENITEZ2018,
title = {A simple and efficient numerical procedure to compute the inverse Langevin function with high accuracy},
journal = {Journal of Non-Newtonian Fluid Mechanics},
volume = {261},
pages = {153-163},
year = {2018},
issn = {0377-0257},
doi = {https://doi.org/10.1016/j.jnnfm.2018.08.011},
author = {José María Benítez and Francisco Javier Montáns},
}

@article{MUSIKHIN2023,
title = {To the theory of magneto-induced flow in thrombosed channels},
journal = {Journal of Magnetism and Magnetic Materials},
volume = {587},
pages = {171316},
year = {2023},
issn = {0304-8853},
doi = {https://doi.org/10.1016/j.jmmm.2023.171316},
url = {https://www.sciencedirect.com/science/article/pii/S0304885323009666},
author = {A. Yu. Musikhin and P. Kuzhir and A. Yu. Zubarev}
}

@book{skomskimagnetism2008,
  title     = {Simple Models of Magnetism},
  author    = {Skomski, Ralph},
  year      = {2008},
  publisher = {Oxford University Press},
  address   = {Oxford},
  isbn      = {9780198570752}
}

@article{rosenberg2023influence,
  title={The influence of anisotropy on the microstructure and magnetic properties of dipolar nanoplatelet suspensions},
  author={Rosenberg, Margaret and Kantorovich, Sofia},
  journal={Phys. Chem. Chem. Phys.},
  year={2023},
  volume={25},
  pages={2781--2792},
  doi={10.1039/D2CP03360G},
  publisher={Royal Society of Chemistry}
}

@article{butcher2000numerical,
title = {Numerical methods for ordinary differential equations in the 20th century},
journal = {Journal of Computational and Applied Mathematics},
volume = {125},
number = {1},
pages = {1-29},
year = {2000},
note = {Numerical Analysis 2000. Vol. VI: Ordinary Differential Equations and Integral Equations},
issn = {0377-0427},
doi = {https://doi.org/10.1016/S0377-0427(00)00455-6},
url = {https://www.sciencedirect.com/science/article/pii/S0377042700004556},
author = {J.C. Butcher}
}

@Article{2013-kantorovich-sm,
  Title                    = {The influence of shape anisotropy on the microstructure of magnetic dipolar particles},
  Author                   = {Kantorovich, Sofia and Pyanzina, Elena and Sciortino, Francesco},
  Journal                  = {Soft Matter},
  Year                     = {2013},
  Pages                    = {6594--6603},
  Volume                   = {9},
  Doi                      = {10.1039/C3SM50197C}
}

@article{Khelfallah2026,
    author = {Khelfallah, Malika and Novak, Ekaterina V. and Kuznetsov, Andrey A. and Mostarac, Deniz and Daffé, Niéli and Sikora, Marcin and Neveu, Sophie and Zečević, Jovana and Meeldijk, Johannes D. and Taverna, Dario and Sainctavit, Philippe and Rovezzi, Mauro and Elnaggar, Hebatalla and Bertuit, Enzo and Mille, Nicolas and Belkhou, Rachid and Dupuis, Vincent and Carvallo, Claire and Juhin, Amélie and Kantorovich, Sofia S.},
    title = {Hard meets soft: tuning binary ferrofluids},
    journal = {Nanoscale},
    volume = {18},
    number = {22},
    pages = {11724-11738},
    year = {2026},
    month = {06},
    issn = {2040-3364},
    doi = {10.1039/d5nr05218a}
}

@Article{steinbach16a,
  Title                    = {Bistable self-assembly in homogeneous colloidal systems for flexible modular architectures},
  Author                   = {Steinbach, G. and Nissen, D. and Albrecht, M. and Novak, E. V. and S\'anchez, P. A. and Kantorovich, S. S. and Gemming, S. and Erbe, A.},
  Journal                  = {Soft Matter},
  Year                     = {2016},
  Pages                    = {2737-2743},
  Volume                   = {12},
  Doi                      = {10.1039/C5SM02899J},
  Issue                    = {10},
  Publisher                = {The Royal Society of Chemistry}
}

@article{Krekhov2017,
  title = {Spontaneous Core Rotation in Ferrofluid Pipe Flow},
  author = {Krekhov, Alexei and Shliomis, Mark},
  journal = {Phys. Rev. Lett.},
  volume = {118},
  issue = {11},
  pages = {114503},
  numpages = {5},
  year = {2017},
  month = {Mar},
  publisher = {American Physical Society},
  doi = {10.1103/PhysRevLett.118.114503},
  url = {https://link.aps.org/doi/10.1103/PhysRevLett.118.114503}
}

@book{BlumsCebersMaiorov1997,
title = {Magnetic Fluids},
author = {Elmars Blums and Andrejs Cebers and M. M. Maiorov},
publisher = {De Gruyter},
address = {Berlin, New York},
doi = {doi:10.1515/9783110807356},
isbn = {9783110807356},
year = {1997},
lastchecked = {2026-02-17}
}

@article{Shliomis2001,
  title = {Ferrohydrodynamics: Testing a third magnetization equation},
  author = {Shliomis, Mark I.},
  journal = {Phys. Rev. E},
  volume = {64},
  issue = {6},
  pages = {060501},
  numpages = {4},
  year = {2001},
  month = {Nov},
  publisher = {American Physical Society},
  doi = {10.1103/PhysRevE.64.060501},
}

@article{wolfschwenger2024molecular,
  title={Molecular dynamics modelling of interacting magnetic nanoparticles for investigating equilibrium and dynamic ensemble properties},
  author={Wolfschwenger, Manuel and Jaufenthaler, Aaron and Hanser, Friedrich and Gamper, Jakob and Hofer, Thomas S and Baumgarten, Daniel},
  journal={Applied Mathematical Modelling},
  volume={136},
  pages={115624},
  year={2024},
  publisher={Elsevier}
}

@article{Fang2022,
doi = {10.1088/1361-648X/ac4345},
year = {2021},
month = {dec},
publisher = {IOP Publishing},
volume = {34},
number = {11},
pages = {115102},
author = {Fang, Angbo},
title = {Dynamical effective field model for interacting ferrofluids: I. Derivations for homogeneous, inhomogeneous, and polydisperse cases},
journal = {Journal of Physics: Condensed Matter},
}

@article{Kuznetsov2022,
title = {Nonlinear response of a dilute ferrofluid to an alternating magnetic field},
journal = {Journal of Molecular Liquids},
volume = {346},
pages = {117449},
year = {2022},
issn = {0167-7322},
doi = {https://doi.org/10.1016/j.molliq.2021.117449},
author = {Andrey A. Kuznetsov and Alexander F. Pshenichnikov},
}

@article{Rinaldi2010,
  title = {Magnetoviscosity in dilute ferrofluids from rotational Brownian dynamics simulations},
  author = {Soto-Aquino, D. and Rinaldi, C.},
  journal = {Phys. Rev. E},
  volume = {82},
  issue = {4},
  pages = {046310},
  numpages = {10},
  year = {2010},
  month = {Oct},
  publisher = {American Physical Society},
  doi = {10.1103/PhysRevE.82.046310},
}

@article{martsenyuk1974kinetics,
  author       = {Martsenyuk, M. A. and Raikher, Yu. L. and Shliomis, M. I.},
  title        = {On the kinetics of magnetization of suspensions of ferromagnetic particles},
  journal      = {Sov. Phys. JETP},
  year         = {1974},
  volume       = {38},
  number       = {2},
  pages        = {413--416},
  publisher    = {Pleiades Publishing, Ltd.(Плеадес Паблишинг, Лтд)}
}

@article{raikher1994effective,
  author       = {Raikher, Yuriy L. and Shliomis, Mark I.},
  title        = {The effective field method in the orientational kinetics of magnetic fluids},
  journal      = {Adv. Chem. Phys.},
  year         = {1994},
  volume       = {87},
  pages        = {595--751},
  doi          = {10.1002/9780470141465.ch8},
}

@article{Rusakov2021,
	author = {Rusakov, V. V. and Raikher, Yu. L.},
	doi = {10.1134/S1061933X21010117},
	isbn = {1608-3067},
	journal = {Colloid Journal},
	number = {1},
	pages = {116--126},
	title = {Nonlinear Magnetic Response of a Viscoelastic Ferrocolloid: Effective Field Approximation},
	volume = {83},
	year = {2021},
}

@article{Tripathi2025,
	author = {Tripathi, Ashwani Kr. and Morozov, Konstantin I. and Rubinstein, Boris Y. and Leshansky, Alexander M.},
	doi = {10.1038/s42005-025-02379-5},
	id = {Tripathi2025},
	isbn = {2399-3650},
	journal = {Communications Physics},
	number = {1},
	pages = {476},
	title = {Weak thermal fluctuations impede steering of chiral magnetic nanobots},
	volume = {8},
	year = {2025},
}

@article{poperechny2023multipeak,
  title={Multipeak dynamic magnetic susceptibility of a superparamagnetic nanoparticle suspended in a fluid},
  author={Poperechny, IS},
  journal={Phys. Rev. B},
  volume={107},
  number={6},
  pages={064416},
  year={2023},
  publisher={APS},
  doi={10.1103/PhysRevB.107.064416}
}

@article{taukulis2012coupled,
  title={Coupled stochastic dynamics of magnetic moment and anisotropy axis of a magnetic nanoparticle},
  author={Taukulis, R and Cebers, A},
  journal={Phys. Rev. E},
  volume={86},
  number={6},
  pages={061405},
  year={2012},
  publisher={APS},
  doi={10.1103/PhysRevE.86.061405}
}

@article{kim2010biofunctionalized,
  title={Biofunctionalized magnetic-vortex microdiscs for targeted cancer-cell destruction},
  author={Kim, Dong-Hyun and Rozhkova, Elena A and Ulasov, Ilya V and Bader, Samuel D and Rajh, Tijana and Lesniak, Maciej S and Novosad, Valentyn},
  journal={Nature materials},
  volume={9},
  number={2},
  pages={165--171},
  year={2010},
  publisher={Nature Publishing Group UK London}
}

@article{welbourne2021high,
  title={High-yield fabrication of perpendicularly magnetised synthetic antiferromagnetic nanodiscs},
  author={Welbourne, Emma N and Vemulkar, Tarun and Cowburn, Russell P},
  journal={Nano Research},
  volume={14},
  number={11},
  pages={3873--3878},
  year={2021},
  publisher={Springer}
}

@Article{D5BM00739A,
author ="Scheibler, S. and Wei, H. and Ackers, J. and Helbig, S. and Koraltan, S. and Peremadathil-Pradeep, R. and Krupinski, M. and Graeser, M. and Suess, D. and Herrmann, I. K. and Hug, H. J.",
title  ="Approaching the physical limits of specific absorption rate for synthetic antiferromagnetic nanodisks in hyperthermia applications",
journal  ="Biomater. Sci.",
year  ="2025",
volume  ="13",
issue  ="22",
pages  ="6285-6297",
publisher  ="The Royal Society of Chemistry",
doi  ="10.1039/D5BM00739A",
url  ="http://dx.doi.org/10.1039/D5BM00739A"}

@article{serantes2021nanoparticle,
  title={Nanoparticle size threshold for magnetic agglomeration and associated hyperthermia performance},
  author={Serantes, David and Baldomir, Daniel},
  journal={Nanomaterials},
  volume={11},
  number={11},
  pages={2786},
  year={2021},
  publisher={MDPI},
  doi = {10.3390/nano11112786}
}

@article{carrey2011simple,
  title={Simple models for dynamic hysteresis loop calculations of magnetic single-domain nanoparticles: Application to magnetic hyperthermia optimization},
  author={Carrey, Julian and Mehdaoui, Boubker and Respaud, Marc},
  journal={Journal of applied physics},
  volume={109},
  number={8},
  year={2011},
  publisher={AIP Publishing}
}

@article{felderhof_hydrodynamics_1999,
	author = {Felderhof, B. U. and Kroh, H. J.},
	doi = {10.1063/1.478642},
	issn = {0021-9606, 1089-7690},
	journal = {The Journal of Chemical Physics},
	language = {en},
	month = apr,
	number = {15},
	pages = {7403--7411},
	title = {Hydrodynamics of magnetic and dielectric fluids in interaction with the electromagnetic field},
	url = {http://aip.scitation.org/doi/10.1063/1.478642},
	urldate = {2022-06-15},
	volume = {110},
	year = {1999}}

@article{muller_structure_2001,
	author = {M{\"u}ller, H W and Liu, M},
	journal = {Physical review E, Statistical, nonlinear, and soft matter physics},
	number = {6 Pt 1},
	pages = {061405},
	title = {Structure of ferrofluid dynamics},
	url = {http://pre.aps.org/abstract/PRE/v64/i6/e061405},
	volume = {64},
	year = {2001}}

@incollection{shliomis_theory_1994,
	address = {Hoboken, NJ, USA},
	author = {Shliomis, M I and Stepanov, V I},
	booktitle = {Advances in {Chemical} {Physics}},
	doi = {10.1002/9780470141465.ch1},
	isbn = {978-0-471-30312-1},
	month = jan,
	note = {Journal Abbreviation: Advances in Chemical Physics},
	pages = {1--30},
	publisher = {John Wiley \& Sons, Inc.},
	title = {Theory of the {Dynamic} {Susceptibility} of {Magnetic} {Fluids}},
	url = {http://doi.wiley.com/10.1002/9780470141465.ch1},
	volume = {87},
	year = {1994}}

@article{bean_superparamagnetism_1959,
	author = {Bean, C. P. and Livingston, J. D.},
	doi = {10.1063/1.2185850},
	issn = {0021-8979, 1089-7550},
	journal = {Journal of Applied Physics},
	language = {en},
	month = apr,
	number = {4},
	pages = {S120--S129},
	title = {Superparamagnetism},
	url = {https://pubs.aip.org/jap/article/30/4/S120/388906/Superparamagnetism},
	urldate = {2025-12-30},
	volume = {30},
	year = {1959}}

@article{Gilbert1955,
	author = {Thomas L. Gilbert},
	journal = {Physical Review D},
	pages = {1243},
	title = {A Lagrangian Formulation of the Gyromagnetic Equation of the Magnetization Field},
	url = {https://api.semanticscholar.org/CorpusID:197507556},
	volume = {100},
	year = {1955}}

@article{landau_theory_1935,
	author = {Landau, L and Lifshits, E},
	journal = {Phys. Zeitsch. der Sow.},
	language = {en},
	pages = {153--169},
	title = {{ON} {THE} {THEORY} {OF} {THE} {DISPERSION} {OF} {MAGNETIC} {PERMEABILITY} {IN} {FERROMAGNETIC} {BODIES}},
	volume = {8},
	year = {1935}}

@article{kubo_brownian_1970,
	author = {Kubo, Ryogo and Hashitsume, Natsuki},
	doi = {10.1143/PTPS.46.210},
	issn = {0375-9687},
	journal = {Progress of Theoretical Physics Supplement},
	language = {en},
	pages = {210--220},
	title = {Brownian {Motion} of {Spins}},
	url = {https://academic.oup.com/ptps/article-lookup/doi/10.1143/PTPS.46.210},
	urldate = {2025-12-30},
	volume = {46},
	year = {1970}}

@article{PYANZINA2025,
title = {Dynamic magnetic response of multicore particles: The role of grain magnetic anisotropy and intergrain interactions},
journal = {Journal of Molecular Liquids},
volume = {421},
pages = {126842},
year = {2025},
issn = {0167-7322},
doi = {https://doi.org/10.1016/j.molliq.2024.126842},
url = {https://www.sciencedirect.com/science/article/pii/S0167732224029040},
author = {Elena S. Pyanzina and Ekaterina V. Novak and Andrey A. Kuznetsov and Sofia S. Kantorovich}
}

@article{garcia-palacios_langevin-dynamics_1998,
	author = {Garc{\'\i}a-Palacios, Jos{\'e} Luis and L{\'a}zaro, Francisco J},
	doi = {10.1103/PhysRevB.58.14937},
	journal = {Physical Review B},
	note = {Publisher: American Physical Society},
	number = {22},
	pages = {14937--14958},
	title = {Langevin-dynamics study of the dynamical properties of small magnetic particles},
	url = {http://link.aps.org/doi/10.1103/PhysRevB.58.14937},
	volume = {58},
	year = {1998}}

@article{iglesias_fundamental_2022,
	author = {Iglesias, C. A. and De Ara{\'u}jo, J. C. R. and Silva, E. F. and Gamino, M. and Correa, M. A. and Bohn, F.},
	doi = {10.1103/PhysRevB.106.094405},
	issn = {2469-9950, 2469-9969},
	journal = {Physical Review B},
	language = {en},
	month = sep,
	number = {9},
	pages = {094405},
	title = {Fundamental inequalities in the {Stoner}-{Wohlfarth} model},
	url = {https://link.aps.org/doi/10.1103/PhysRevB.106.094405},
	urldate = {2023-08-12},
	volume = {106},
	year = {2022}}

@article{lu_field_1994,
	author = {Lu, Jing Ju and Huang, Huei Li and Klik, I.},
	doi = {10.1063/1.358424},
	issn = {0021-8979, 1089-7550},
	journal = {Journal of Applied Physics},
	language = {en},
	month = aug,
	number = {3},
	pages = {1726--1732},
	title = {Field orientations and sweep rate effects on magnetic switching of {Stoner}--{Wohlfarth} particles},
	url = {https://pubs.aip.org/jap/article/76/3/1726/528206/Field-orientations-and-sweep-rate-effects-on},
	urldate = {2024-04-25},
	volume = {76},
	year = {1994}}

@article{usov_hysteresis_2009,
	author = {Usov, N. A. and Grebenshchikov, Yu. B.},
	doi = {10.1063/1.3173280},
	issn = {0021-8979, 1089-7550},
	journal = {Journal of Applied Physics},
	language = {en},
	month = jul,
	number = {2},
	pages = {023917},
	title = {Hysteresis loops of an assembly of superparamagnetic nanoparticles with uniaxial anisotropy},
	url = {https://pubs.aip.org/jap/article/106/2/023917/393635/Hysteresis-loops-of-an-assembly-of},
	urldate = {2024-04-25},
	volume = {106},
	year = {2009}}

@article{chuev_nanomagnetism_2007,
	author = {Chuev, M A and Hesse, J},
	doi = {10.1088/0953-8984/19/50/506201},
	issn = {0953-8984, 1361-648X},
	journal = {Journal of Physics: Condensed Matter},
	language = {en},
	month = dec,
	number = {50},
	pages = {506201},
	shorttitle = {Nanomagnetism},
	title = {Nanomagnetism: extension of the {Stoner}--{Wohlfarth} model within {N{\'e}el}'s ideas and useful plots},
	url = {https://iopscience.iop.org/article/10.1088/0953-8984/19/50/506201},
	urldate = {2024-11-26},
	volume = {19},
	year = {2007}}

@article{stoner_mechanism_1948,
	author = {Stoner, E C and Wohlfarth, E P},
	journal = {Philosophical Transactions of the Royal Society A: Mathematical, Physical and Engineering Sciences},
	language = {en},
	number = {826},
	pages = {599--642},
	title = {A mechanism of magnetic hysteresis in heterogeneous alloys},
	volume = {240},
	year = {1948}}

@article{usadel_dynamics_2015,
        author = {Usadel, K. D. and Usadel, C.},
        doi = {10.1063/1.4937919},
        issn = {0021-8979, 1089-7550},
        journal = {Journal of Applied Physics},
        language = {en},
        month = dec,
        number = {23},
        pages = {234303},
        title = {Dynamics of magnetic single domain particles embedded in a viscous liquid},
        url = {http://aip.scitation.org/doi/10.1063/1.4937919},
        urldate = {2022-06-01},
        volume = {118},
        year = {2015}}

@article{usadel_dynamics_2017,
        author = {Usadel, Klaus D.},
        doi = {10.1103/PhysRevB.95.104430},
        issn = {2469-9950, 2469-9969},
        journal = {Physical Review B},
        language = {en},
        month = mar,
        number = {10},
        pages = {104430},
        title = {Dynamics of magnetic nanoparticles in a viscous fluid driven by rotating magnetic fields},
        url = {https://link.aps.org/doi/10.1103/PhysRevB.95.104430},
        urldate = {2022-06-01},
        volume = {95},
        year = {2017}}

@article{kroger_combined_2022,
        author = {Kr{\"o}ger, Martin and Ilg, Patrick},
        doi = {10.1142/S0218202522500300},
        issn = {0218-2025, 1793-6314},
        journal = {Mathematical Models and Methods in Applied Sciences},
        language = {en},
        month = jun,
        pages = {1349--1383},
        shorttitle = {Combined dynamics of magnetization and particle rotation of a suspended superparamagnetic particle in the presence of an orienting field},
        title = {Combined dynamics of magnetization and particle rotation of a suspended superparamagnetic particle in the presence of an orienting field: {Semi}-analytical and numerical solution},
        url = {https://www.worldscientific.com/doi/10.1142/S0218202522500300},
        urldate = {2022-06-23},
        volume = {32},
        year = {2022}}

@article{poperechny_combined_2020,
        author = {Poperechny, I.S.},
        doi = {10.1016/j.molliq.2019.112109},
        issn = {01677322},
        journal = {Journal of Molecular Liquids},
        language = {en},
        month = feb,
        pages = {112109},
        shorttitle = {Combined rotational diffusion of a superparamagnetic particle and its magnetic moment},
        title = {Combined rotational diffusion of a superparamagnetic particle and its magnetic moment: {Solution} of the kinetic equation},
        url = {https://linkinghub.elsevier.com/retrieve/pii/S0167732219355679},
        urldate = {2021-06-09},
        volume = {299},
        year = {2020}}

@article{coffey_thermal_2012,
        title = {Thermal fluctuations of magnetic nanoparticles: {Fifty} years after {Brown}},
        volume = {112},
        url = {http://aip.scitation.org/doi/10.1063/1.4754272},
        doi = {10.1063/1.4754272},
        number = {12},
        journal = {Journal Of Applied Physics},
        author = {Coffey, William T and Kalmykov, Yuri P},
        year = {2012},
        note = {Publisher: AIP Publishing LLC},
        pages = {121301--99},
}

@article{Wernsdorfer:1997vk,
        author = {Wernsdorfer, W and Orozcho, E Bonet and Hasselbach, K and Benoit, A and Barbara, B and Demoncy, N and Loiseau, A and Pascard, H and Mailly, D},
        journal = {Physical Review Letters},
        month = mar,
        number = {9},
        pages = {1791--1794},
        title = {{Experimental Evidence of the N{\'e}el-Brown Model of Magnetization Reversal}},
        volume = {78},
        year = {1997}}

@article{keshtgar_magnetomechanical_2017,
        title={Magnetomechanical coupling and ferromagnetic resonance in magnetic nanoparticles},
        author={Keshtgar, Hedyeh and Streib, Simon and Kamra, Akashdeep and Blanter, Yaroslav M and Bauer, Gerrit E W},
        journal={Physical Review B},
        volume={95},
        number={13},
        pages={134447},
        year={2017},
        publisher={APS}
}

@article{helbig_self-consistent_2023,
	title = {Self-consistent solution of magnetic and friction energy losses of a magnetic nanoparticle},
	volume = {107},
	url = {https://link.aps.org/doi/10.1103/PhysRevB.107.054416},
	doi = {10.1103/PhysRevB.107.054416},
	number = {5},
	urldate = {2025-05-21},
	journal = {Physical Review B},
	author = {Helbig, Santiago and Abert, Claas and Sánchez, Pedro A. and Kantorovich, Sofia S. and Suess, Dieter},
	month = feb,
	year = {2023},
	note = {Publisher: American Physical Society},
	pages = {054416},
}

@article{durhuus_conservation_2024,
	title = {Conservation laws for interacting magnetic nanoparticles at finite temperature},
	volume = {109},
	url = {https://link.aps.org/doi/10.1103/PhysRevB.109.054421},
	doi = {10.1103/PhysRevB.109.054421},
	number = {5},
	urldate = {2025-12-17},
	journal = {Physical Review B},
	author = {Durhuus, Frederik L. and Beleggia, Marco and Frandsen, Cathrine},
	month = feb,
	year = {2024},
	note = {Publisher: American Physical Society},
	pages = {054421},
}

@article{barnett_magnetization_1915,
	title = {Magnetization by {Rotation}},
	volume = {6},
	url = {https://link.aps.org/doi/10.1103/PhysRev.6.239},
	doi = {10.1103/PhysRev.6.239},
	number = {4},
	urldate = {2025-10-14},
	journal = {Physical Review},
	author = {Barnett, S. J.},
	month = oct,
	year = {1915},
	note = {Publisher: American Physical Society},
	pages = {239--270},
}

@article{einstein_experimental_1915,
	title = {Experimental proof of the existence of Amp{\`e}re's molecular currents},
	author = {Einstein, Albert and de Haas, Wander Johannes},
	journal = {Proceedings of the Royal Netherlands Academy of Arts and Sciences},
	volume = {18},
	pages = {696--711},
	year = {1915}
}

@article{martinez_micromagnetic_2007,
	series = {Proceedings of the {Joint} {European} {Magnetic} {Symposia}},
	title = {Micromagnetic simulations with thermal noise: {Physical} and numerical aspects},
	volume = {316},
	issn = {0304-8853},
	shorttitle = {Micromagnetic simulations with thermal noise},
	url = {https://www.sciencedirect.com/science/article/pii/S0304885307005276},
	doi = {10.1016/j.jmmm.2007.03.178},
	number = {2},
	urldate = {2026-01-16},
	journal = {Journal of Magnetism and Magnetic Materials},
	author = {Martinez, E. and Lopez-Diaz, L. and Torres, L. and Garcia-Cervera, C. J.},
	month = sep,
	year = {2007},
	pages = {269--272},
}

@mastersthesis{zehner_integrating_2024,
	author      = {Zehner, David},
	title       = {Integrating molecular dynamics and micromagnetics for the investigation of nanoparticles in liquid carriers},
	school = {University of Vienna},
	type        = {Master’s thesis},
	year        = {2024},
	location    = {Vienna, Austria},
	url         = {https://doi.org/10.25365/thesis.76830},
	doi         = {10.25365/thesis.76830},
}

@article{bruckner_magnumnp_2023,
	title = {magnum.np: a {PyTorch} based {GPU} enhanced finite difference micromagnetic simulation framework for high level development and inverse design},
	volume = {13},
	copyright = {2023 The Author(s)},
	issn = {2045-2322},
	shorttitle = {magnum.np},
	url = {https://www.nature.com/articles/s41598-023-39192-5},
	doi = {10.1038/s41598-023-39192-5},
	language = {en},
	number = {1},
	urldate = {2024-04-22},
	journal = {Scientific Reports},
	publisher = {Nature Publishing Group},
	author = {Bruckner, Florian and Koraltan, Sabri and Abert, Claas and Suess, Dieter},
	month = jul,
	year = {2023},
	pages = {12054},
}

@article{Satoh99,
        author = {A. Satoh and R. W. Chantrell and G. N. Coverdale},
        journal = {J. Coll. Interf. Sci.},
        pages = {44-59},
        title = {Brownian Dynamics Simulations of ferromagnetic colloidal dispersions in a simple shear flow},
        volume = 209,
        year = 1999}

@article{cichocki_friction_1994,
        title = {Friction and mobility of many spheres in {Stokes} flow},
        volume = {100},
        url = {http://scitation.aip.org/content/aip/journal/jcp/100/5/10.1063/1.466366},
        doi = {10.1063/1.466366},
        number = {5},
        journal = {Journal of Chemical Physics},
        author = {Cichocki, B and Felderhof, B U and Hinsen, K and Wajnryb, E and awzdziewicz, J Bl},
        year = {1994},
        note = {Publisher: AIP Publishing},
        pages = {3780--3790},
}

@book{allenbook,
        address = {Oxford},
        author = {M. P. Allen and D. J. Tildesley},
        publisher = {Oxford University Press},
        title = {Computer Simulation of Liquids},
        year = {1987}
}

@book{Dhont_book,
        address = {Amsterdam},
        author = {J. K. G. Dhont},
        publisher = {Elsevier},
        series = {Studies in {I}nterface {S}cience},
        title = {An introduction to dynamics of colloids},
        year = {1996}
}

@incollection{Ilg_lnp,
        address = {Berlin},
        author = {P. Ilg and S. Odenbach},
        booktitle = {Colloidal Magnetic Fluids: Basics, Development and Applications of Ferrofluids},
        chapter = {Ferrofluid structure and rheology},
        editor = {S. Odenbach},
        pages = {245-315},
        publisher = {Springer},
        series = {Lecture Notes in Physics},
        title = {Ferrofluid structure and rheology},
        volume = {763},
        year = {2008}
}

@article{Sreekumari2013,
        author = {A. Sreekumari and P. Ilg},
        issue = {4},
        journal = {Physical Review E},
        pages = {042315},
        title = {Slow relaxation in structure-forming ferrofluids},
        volume = {88},
        year = {2013}
}

@article{Sreekumari2015,
        author = {Sreekumari, Aparna and Ilg, Patrick},
        journal = {Physical Review E},
        month = jul,
        number = {1},
        pages = {012306},
        title = {{Anisotropy of magnetoviscous effect in structure-forming ferrofluids}},
        volume = {92},
        year = {2015}
}

@article{cerda_aggregate_2008,
        title = {Aggregate formation in ferrofluid monolayers: simulations and theory},
        volume = {20},
        url = {http://www.iop.org/EJ/abstract/0953-8984/20/20/204125/},
        doi = {10.1088/0953-8984/20/20/204125},
        number = {20},
        journal = {Journal of Physics: Condensed Matter},
        author = {Cerda, Juan J and Kantorovich, Sofia and Holm, Christian},
        month = jan,
        year = {2008},
        pages = {204125}
}

@article{ilg_magnetoviscosity_2005,
        title = {Magnetoviscosity of semidilute ferrofluids and the role of dipolar interactions: {Comparison} of molecular simulations and dynamical mean-field theory},
        volume = {71},
        url = {http://link.aps.org/doi/10.1103/PhysRevE.71.031205},
        doi = {10.1103/PhysRevE.71.031205},
        number = {3},
        journal = {Physical review E, Statistical, nonlinear, and soft matter physics},
        author = {Ilg, P and Kröger, M and Hess, S},
        year = {2005}
}

@article{jordanovic_structure_2009,
        title = {Structure of ferrofluid nanofilms in homogeneous magnetic fields},
        volume = {79},
        url = {http://pre.aps.org/abstract/PRE/v79/i2/e021405},
        doi = {10.1103/PhysRevE.79.021405},
        number = {2},
        journal = {Physical Review E},
        author = {Jordanovic, Jelena and Klapp, Sabine H L},
        month = jan,
        year = {2009},
        pages = {021405}
}

@article{rosa_shear_2020,
        title = {Shear rate dependence of viscosity and normal stress differences in ferrofluids},
        volume = {499},
        issn = {03048853},
        url = {https://linkinghub.elsevier.com/retrieve/pii/S0304885319327702},
        doi = {10.1016/j.jmmm.2019.166184},
        language = {en},
        urldate = {2024-09-11},
        journal = {Journal of Magnetism and Magnetic Materials},
        author = {Rosa, Adriano P. and Cunha, Francisco R.},
        month = apr,
        year = {2020},
        pages = {166184}
}

@article{zverev_computer_2021,
        title = {Computer {Simulations} of {Dynamic} {Response} of {Ferrofluids} on an {Alternating} {Magnetic} {Field} with {High} {Amplitude}},
        volume = {9},
        issn = {2227-7390},
        url = {https://www.mdpi.com/2227-7390/9/20/2581},
        doi = {10.3390/math9202581},
        language = {en},
        number = {20},
        urldate = {2023-05-17},
        journal = {Mathematics},
        author = {Zverev, Vladimir and Dobroserdova, Alla and Kuznetsov, Andrey and Ivanov, Alexey and Elfimova, Ekaterina},
        month = oct,
        year = {2021},
        pages = {2581}
}

@article{pi_hex,
        author = {P. Ilg},
        journal = {Eur. Phys. J. E},
        pages = {169-176},
        title = {Importance of depletion interactions for structure and dynamics of ferrofluids},
        volume = {26},
        year = {2008}
}

@article{Morimoto2002,
        author = {H. Morimoto and T. Maekawa and Y. Matsumo},
        journal = {Physical Review E},
        pages = {061508},
        title = {Nonequilibrium Brownian dynamics analysis of negative viscosity induced in a magnetic fluid subjected to both ac magnetic and shear flow fields},
        volume = {65},
        year = {2002}
}

@article{ilg_structure_2006,
        title = {Structure and rheology of ferrofluids: simulation results and kinetic models},
        volume = {18},
        url = {http://stacks.iop.org/0953-8984/18/i=38/a=S15?key=crossref.c4282d50ea142d2477ba365bf90343d8},
        doi = {10.1088/0953-8984/18/38/S15},
        number = {38},
        journal = {Journal of Physics: Condensed Matter},
        author = {Ilg, Patrick and Coquelle, Eric and Hess, Siegfried},
        month = jan,
        year = {2006},
        pages = {S2757--S2770}
}

@article{zablotsky_field_2019,
        author = {Zablotsky, Dmitry},
        date = {2019-03},
        doi = {10.1016/j.jmmm.2018.10.065},
        issn = {03048853},
        journal = {Journal of Magnetism and Magnetic Materials},
        journaltitle = {Journal of Magnetism and Magnetic Materials},
        langid = {english},
        pages = {462--466},
        shortjournal = {Journal of Magnetism and Magnetic Materials},
        title = {Field effect in the viscosity of magnetic colloids studied by multi-particle collision dynamics},
        url = {https://linkinghub.elsevier.com/retrieve/pii/S0304885318320134},
        urldate = {2022-11-17},
        volume = {474},
        year = {2019}
}

@article{camp_how_2021,
        author = {Camp, Philip J. and Ivanov, Alexey O. and Sindt, Julien O.},
        date = {2021-06-21},
        doi = {10.1103/PhysRevE.103.062611},
        issn = {2470-0045, 2470-0053},
        journal = {Physical Review E},
        journaltitle = {Physical Review E},
        langid = {english},
        number = {6},
        pages = {062611},
        shortjournal = {Phys. Rev. E},
        title = {How chains and rings affect the dynamic magnetic susceptibility of a highly clustered ferrofluid},
        url = {https://link.aps.org/doi/10.1103/PhysRevE.103.062611},
        urldate = {2023-09-10},
        volume = {103},
        year = {2021}
}

@article{ivanov_effects_2022,
        author = {Ivanov, Alexey O. and Camp, Philip J.},
        date = {2022-06},
        doi = {10.1016/j.molliq.2022.119034},
        issn = {01677322},
        journal = {Journal of Molecular Liquids},
        journaltitle = {Journal of Molecular Liquids},
        langid = {english},
        pages = {119034},
        shortjournal = {Journal of Molecular Liquids},
        title = {Effects of interactions, structure formation, and polydispersity on the dynamic magnetic susceptibility and magnetic relaxation of ferrofluids},
        url = {https://linkinghub.elsevier.com/retrieve/pii/S0167732222005724},
        urldate = {2024-06-03},
        volume = {356},
        year = {2022}
}

@book{rapaport_art_2004,
	address = {Cambridge},
	edition = {2},
	title = {The {Art} of {Molecular} {Dynamics} {Simulation}},
	isbn = {978-0-521-82568-9},
	url = {https://www.cambridge.org/core/books/art-of-molecular-dynamics-simulation/57D40C5ECE9B7EA17C0E77E7754F5874},
	doi = {10.1017/CBO9780511816581},
	urldate = {2025-11-24},
	publisher = {Cambridge University Press},
	author = {Rapaport, D. C.},
	year = {2004},
}

@article{Ali17,
  author  = {Ali, I. and Peng, C. and Naz, I. and Khan, Z. M. and Sultan, M. and Islam, T. and Abbasi, I. A.},
  title   = {Phytogenic magnetic nanoparticles for wastewater treatment: a review},
  journal = {RSC Advances},
  volume  = {7},
  number  = {64},
  pages   = {40158--40178},
  year    = {2017}
}

@article{Ars23,
  author  = {Arsalani, S. and Radon, P. and Eberbeck, D. and K{\"o}rber, R. and Jaufenthaler, A. and Baumgarten, D. and Wiekhorst, F.},
  title   = {Temperature dependent magnetorelaxometry of magnetic nanoparticle ensembles},
  journal = {Physics in Medicine and Biology},
  volume  = {68},
  number  = {17},
  pages   = {175017},
  year    = {2023}
}

@article{Coe22,
  author  = {Coene, A. and Leliaert, J.},
  title   = {Magnetic nanoparticles in theranostic applications},
  journal = {Journal of Applied Physics},
  volume  = {131},
  number  = {16},
  pages   = {160902},
  year    = {2022}
}

@article{For18,
  author  = {Forouzandehmehr, M. and Shamloo, A.},
  title   = {Margination and adhesion of micro- and nanoparticles in the coronary circulation: a step towards optimised drug carrier design},
  journal = {Biomechanics and Modeling in Mechanobiology},
  volume  = {17},
  number  = {1},
  pages   = {205--221},
  year    = {2018}
}

@article{Glo19,
  author  = {Gloag, L. and Mehdipour, M. and Chen, D. and Tilley, R. D. and Gooding, J. J.},
  title   = {Advances in the application of magnetic nanoparticles for sensing},
  journal = {Advanced Materials},
  volume  = {31},
  number  = {48},
  pages   = {1904385},
  year    = {2019}
}

@article{Gon21,
  author  = {Gonella, V. C. and Hanser, F. and Vorwerk, J. and Odenbach, S. and Baumgarten, D.},
  title   = {Influence of local particle concentration gradient forces on the flow-mediated mass transport in a numerical model of magnetic drug targeting},
  journal = {Journal of Magnetism and Magnetic Materials},
  volume  = {525},
  pages   = {167490},
  year    = {2021}
}

@article{Jau24,
  author  = {Jaufenthaler, A. and Sander, T. and Schier, P. and Pansegrau, K. and Wiekhorst, F. and Baumgarten, D.},
  title   = {Human head sized magnetorelaxometry imaging of magnetic nanoparticles with optically pumped magnetometers---A feasibility study},
  journal = {Journal of Magnetism and Magnetic Materials},
  volume  = {596},
  pages   = {171983},
  year    = {2024}
}

@article{Ken16,
  author  = {Kenjere{\v{s}}, S.},
  title   = {On recent progress in modelling and simulations of multi-scale transfer of mass, momentum and particles in bio-medical applications},
  journal = {Flow, Turbulence and Combustion},
  volume  = {96},
  number  = {3},
  pages   = {837--860},
  year    = {2016}
}

@article{Leo15,
  author  = {Leong, S. S. and Ahmad, Z. and Lim, J.},
  title   = {Magnetophoresis of superparamagnetic nanoparticles at low field gradient: hydrodynamic effect},
  journal = {Soft Matter},
  volume  = {11},
  number  = {35},
  pages   = {6968--6980},
  year    = {2015}
}

@article{Lin21,
  author  = {Lindemann, M. C. and Luttke, T. and Nottrodt, N. and Schmitz-Rode, T. and Slabu, I.},
  title   = {FEM based simulation of magnetic drug targeting in a multibranched vessel model},
  journal = {Computer Methods and Programs in Biomedicine},
  volume  = {210},
  pages   = {106354},
  year    = {2021}
}

@article{Maj16,
  author  = {Majidi, S. and Zeinali Sehrig, F. and Samiei, M. and Milani, M. and Abbasi, E. and Dadashzadeh, K. and Akbarzadeh, A.},
  title   = {Magnetic nanoparticles: Applications in gene delivery and gene therapy},
  journal = {Artificial Cells, Nanomedicine, and Biotechnology},
  volume  = {44},
  number  = {4},
  pages   = {1186--1193},
  year    = {2016}
}

@book{Ode02,
  author    = {Odenbach, S. and Thurm, S.},
  title     = {Magnetoviscous Effects in Ferrofluids},
  publisher = {Springer},
  year      = {2002}
}

@article{Ort13,
  author  = {Ortega, D. and Pankhurst, Q. A.},
  title   = {Magnetic hyperthermia},
  journal = {Nanoscience},
  volume  = {1},
  number  = {60},
  pages   = {e88},
  year    = {2013}
}

@article{Pat22,
  author  = {Pathak, S. and Zhang, R. and Gayen, B. and Kumar, V. and Zhang, H. and Pant, R. and Wang, X.},
  title   = {Ultra-low friction self-levitating nanomagnetic fluid bearing for highly efficient wind energy harvesting},
  journal = {Sustainable Energy Technologies and Assessments},
  volume  = {52},
  pages   = {102024},
  year    = {2022}
}

@article{Pri18,
  author  = {Price, P. M. and Mahmoud, W. E. and Al-Ghamdi, A. A. and Bronstein, L. M.},
  title   = {Magnetic drug delivery: where the field is going},
  journal = {Frontiers in Chemistry},
  volume  = {6},
  pages   = {619},
  year    = {2018}
}

@article{Sha19,
  author  = {Shamloo, A. and Amani, A. and Forouzandehmehr, M. and Ghoytasi, I.},
  title   = {In silico study of patient-specific magnetic drug targeting for a coronary LAD atherosclerotic plaque},
  journal = {International Journal of Pharmaceutics},
  volume  = {559},
  pages   = {113--129},
  year    = {2019}
}

@article{Sim18,
  author  = {Simonsen, G. and Strand, M. and {\O}ye, G.},
  title   = {Potential applications of magnetic nanoparticles within separation in the petroleum industry},
  journal = {Journal of Petroleum Science and Engineering},
  volume  = {165},
  pages   = {488--495},
  year    = {2018}
}

@article{2008-cerda-jcp,
	author = {Juan J. Cerd\`{a} and V. Ballenegger and O. Lenz and C. Holm},
	doi = {10.1063/1.3000389},
	journal = {J Chem Phys},
	pages = {234104},
	title = {P3M algorithm for dipolar interactions},
	volume = {129},
	year = {2008}}

@incollection{raikher2004nonlinear,
  title={Nonlinear Dynamic Susceptibilities and Field-Induced Birefringence in Magnetic Particle Assemblies},
  author={Raikher, Yuriy L and Stepanov, Victor I},
  journal={Advances in Chemical Physics},
  volume={129},
  pages={419--588},
  year={2004},
  publisher={Wiley Online Library},
  doi = {10.1002/047168077X.ch4}
}

@article{mcnamara1988use,
  title={Use of the Boltzmann equation to simulate lattice-gas automata},
  author={McNamara, Guy R and Zanetti, Gianluigi},
  journal={Physical review letters},
  volume={61},
  number={20},
  pages={2332},
  year={1988},
  publisher={APS}
}

@article{kruger2017lattice,
  title={The lattice Boltzmann method},
  author={Kr{\"u}ger, Timm and Kusumaatmaja, Halim and Kuzmin, Alexandr and Shardt, Orest and Silva, Goncalo and Viggen, Erlend Magnus},
  journal={Springer International Publishing},
  volume={10},
  number={978-3},
  pages={4--15},
  year={2017},
  publisher={Springer}
}

@article{ilg2019diffusionjump,
    author = {Ilg, Patrick},
    title = {Diffusion-jump model for the combined Brownian and Néel relaxation dynamics of ferrofluids in the presence of external fields and flow},
    journal = {Physical Review E},
    volume = {100}, 
    pages = {022608},
    year = {2019},
    publisher = {APS}
}

@article{ilg2023optimality,
    author = {Ilg, Patrick and Kr\"oger, Martin},
    title = {Field- and concentration-dependent relaxation of magnetic nanoparticles and optimality conditions for magnetic fluid hyperthermia},
    journal = {Scientific Reports},
    volume = {13},
    pages = {16523},
    year = {2023},
    publisher = {Springer}
}

@article{ilg2024nonequilibrium,
    author = {Ilg, Patrick},
    title = {Nonequilibrium response of magnetic nanoparticles to time-varying magnetic fields: Contributions from Brownian and Néel processes},
    journal = {Physical Review E},
    volume = {109},
    pages = {034603},
    year = {2024},
    publisher = {APS}
}

@article{ilg2024stochastic,
    author = {Ilg, Patrick},
    title = {Stochastic thermodynamics and fluctuations in heat released by magnetic nanoparticles in response to time-varying fields},
    journal = {Physical Review B},
    volume = {109},
    pages = {174301},
    year = {2024},
    publisher = {APS}
}

@article{ilg2022longest,
    author = {Ilg, Patrick and Kr\"oger, Martin},
    title = {Longest relaxation time versus maximum loss peak in the field-dependent longitudinal dynamics of suspended magnetic nanoparticles},
    journal = {Physical Review B},
    volume = {106},
    pages = {134433},
    year = {2022},
    publisher = {APS}
}

@article{wang02a,
  title = {Molecular Dynamics Study on the Equilibrium Magnetization Properties and Structure of Ferrofluids},
  author = {Wang, Zuowei and Holm, Christian and M{\"u}ller, Hanns Walter},
  year = 2002,
  month = aug,
  journal = {Physical Review E},
  volume = {66},
  number = {2},
  pages = {021405},
  doi = {10.1103/physreve.66.021405}
}

@article{weeber15,
  title = {Ferrogels Cross-Linked by Magnetic Particles: {{Field-driven}} Deformation and Elasticity Studied Using Computer Simulations},
  author = {Weeber, Rudolf and Kantorovich, Sofia and Holm, Christian},
  year = 2015,
  journal = {The Journal of Chemical Physics},
  volume = {143},
  number = {15},
  pages = {154901},
  doi = {10.1063/1.4932371}
}

@article{brodka04a,
  title = {Ewald Summation Method with Electrostatic Layer Correction for Interactions of Point Dipoles in Slab Geometry},
  author = {Br{\'o}dka, A.},
  year = 2004,
  month = dec,
  journal = {Chemical Physics Letters},
  volume = {400},
  number = {1-3},
  pages = {62--67},
  doi = {10.1016/j.cplett.2004.10.086}
}

@article{weeber19b,
  title = {Accelerating the Calculation of Dipolar Interactions in Particle Based Simulations with Open Boundary Conditions by Means of the {{P}}{$^{2}$}{{NFFT}} Method},
  author = {Weeber, Rudolf and Nestler, Franziska and Weik, Florian and Pippig, Michael and Potts, Daniel and Holm, Christian},
  year = 2019,
  journal = {Journal of Computational Physics},
  volume = {391},
  pages = {243--258},
  issn = {0021-9991},
  doi = {10.1016/j.jcp.2019.01.044},
  e-print = {1808.10341}
}

@article{weik19a,
  title = {{{ESPResSo}} 4.0 -- an Extensible Software Package for Simulating Soft Matter Systems},
  author = {Weik, Florian and Weeber, Rudolf and Szuttor, Kai and Breitsprecher, Konrad and De Graaf, Joost and Kuron, Michael and Landsgesell, Jonas and Menke, Henri and Sean, David and Holm, Christian},
  year = 2019,
  month = mar,
  journal = {The European Physical Journal Special Topics},
  volume = {227},
  number = {14},
  pages = {1789--1816},
  issn = {1951-6355, 1951-6401},
  doi = {10.1140/epjst/e2019-800186-9},
  urldate = {2026-02-17},
  langid = {english}
}

@article{thompson22a,
  title = {{{LAMMPS}} - a Flexible Simulation Tool for Particle-Based Materials Modeling at the Atomic, Meso, and Continuum Scales},
  author = {Thompson, Aidan P. and Aktulga, H. Metin and Berger, Richard and Bolintineanu, Dan S. and Brown, W. Michael and Crozier, Paul S. and In 'T Veld, Pieter J. and Kohlmeyer, Axel and Moore, Stan G. and Nguyen, Trung Dac and Shan, Ray and Stevens, Mark J. and Tranchida, Julien and Trott, Christian and Plimpton, Steven J.},
  year = 2022,
  month = feb,
  journal = {Computer Physics Communications},
  volume = {271},
  pages = {108171},
  publisher = {Elsevier},
  issn = {00104655},
  doi = {10.1016/j.cpc.2021.108171},
  urldate = {2025-03-25},
  langid = {english}
}

@article{ahlrichs99a,
  title = {Simulation of a Single Polymer Chain in Solution by Combining Lattice {{Boltzmann}} and Molecular Dynamics},
  author = {Ahlrichs, Patrick and D{\"u}nweg, Burkhard},
  year = 1999,
  journal = {The Journal of Chemical Physics},
  volume = {111},
  number = {17},
  pages = {8225--8239},
  doi = {10.1063/1.480156}
}

@article{duenweg07a,
  title = {Statistical Mechanics of the Fluctuating Lattice {{Boltzmann}} Equation},
  author = {D{\"u}nweg, Burkhard and Schiller, Ulf D. and Ladd, Anthony J. C.},
  year = 2007,
  month = sep,
  journal = {Physical Review E},
  volume = {76},
  number = {3},
  pages = {036704},
  publisher = {APS},
  issn = {1539-3755, 1550-2376},
  doi = {10.1103/PhysRevE.76.036704},
  urldate = {2026-01-20},
  copyright = {http://link.aps.org/licenses/aps-default-license},
  e-print = {0707.1581},
  langid = {english}
}

@article{novikau26,
  title = {Rheology of {{Magnetic Nanogel Suspensions}}},
  author = {Novikau, Ivan S. and Kantorovich, Sofia S.},
  year = 2026,
  month = jan,
  journal = {The Journal of Physical Chemistry B},
  volume = {130},
  number = {4},
  pages = {1415--1423},
  issn = {1520-6106, 1520-5207},
  doi = {10.1021/acs.jpcb.5c07636},
  urldate = {2026-03-24},
  copyright = {https://doi.org/10.15223/policy-029},
  langid = {english}
}

@article{kim09c,
  title = {Hydrodynamic {{Interactions}} in {{Colloidal Ferrofluids}}: {{A Lattice Boltzmann Study}}},
  shorttitle = {Hydrodynamic {{Interactions}} in {{Colloidal Ferrofluids}}},
  author = {Kim, Eunhye and Stratford, Kevin and Camp, Philip J. and Cates, Michael E.},
  year = 2009,
  month = mar,
  journal = {The Journal of Physical Chemistry B},
  volume = {113},
  number = {12},
  pages = {3681--3693},
  issn = {1520-6106, 1520-5207},
  doi = {10.1021/jp806678m},
  urldate = {2026-03-24},
  langid = {english}
}

@article{kreissl21,
  title = {Frequency-Dependent Magnetic Susceptibility of Magnetic Nanoparticles in a Polymer Solution: A Simulation Study},
  author = {Kreissl, Patrick and Holm, Christian and Weeber, Rudolf},
  year = 2021,
  journal = {Soft Matter},
  volume = {17},
  number = {1},
  pages = {174--183},
  publisher = {The Royal Society of Chemistry},
  doi = {10.1039/D0SM01554G}
}

@article{kreissl23,
  title = {Interplay {{Between Steric}} and {{Hydrodynamic Interactions}} for {{Ellipsoidal Magnetic Nanoparticles}} in a {{Polymer Suspension}}},
  author = {Kreissl, Patrick and Holm, Christian and Weeber, Rudolf},
  year = 2023,
  journal = {Soft Matter},
  volume = {19},
  number = {6},
  pages = {1186--1193},
  publisher = {The Royal Society of Chemistry},
  doi = {10.1039/D2SM01428A}
}

@article{fischer15,
  title = {The Raspberry Model for Hydrodynamic Interactions Revisited. {{I}}. {{Periodic}} Arrays of Spheres and Dumbbells},
  author = {Fischer, Lukas P. and Peter, Toni and Holm, Christian and {de Graaf}, Joost},
  year = 2015,
  journal = {The Journal of Chemical Physics},
  volume = {143},
  number = {8},
  pages = {084107},
  doi = {10.1063/1.4928502}
}

@article{lobaskin04a,
  title = {A New Model for Simulating Colloidal Dynamics},
  author = {Lobaskin, Vladimir and D{\"u}nweg, Burkhard},
  year = 2004,
  month = may,
  journal = {New Journal of Physics},
  volume = {6},
  pages = {54},
  doi = {10.1088/1367-2630/6/1/054}
}

@incollection{arnold13a,
  title = {{{ESPResSo}} 3.1 -- Molecular Dynamics Software for Coarse-Grained Models},
  booktitle = {Meshfree Methods for Partial Differential Equations {{VI}}},
  author = {Arnold, Axel and Lenz, Olaf and Kesselheim, Stefan and Weeber, Rudolf and Fahrenberger, Florian and R{\"o}hm, Dominic and Ko{\v s}ovan, Peter and Holm, Christian},
  editor = {Griebel, Michael and Schweitzer, Marc Alexander},
  year = 2013,
  series = {Lecture Notes in Computational Science and Engineering},
  volume = {89},
  pages = {1--23},
  publisher = {Springer Berlin Heidelberg},
  doi = {10.1007/978-3-642-32979-1_1}
}

@article{ladd94am,
  title = {Numerical Simulations of Particulate Suspensions via a Discretized {{Boltzmann}} Equation. {{Part}} 1. {{Theoretical}} Foundation},
  author = {Ladd, A. J. C.},
  year = 1994,
  journal = {Journal of Fluid Mechanics},
  volume = {271},
  pages = {285--309},
  publisher = {Cambridge Univ Press},
  doi = {10.1017/S0022112094001771}
}

@article{chantrell2000calculations,
  title={Calculations of the susceptibility of interacting superparamagnetic particles},
  author={Chantrell, RW and Walmsley, N and Gore, J and Maylin, M},
  journal={Physical Review B},
  volume={63},
  number={2},
  pages={024410},
  year={2000},
  publisher={APS}
}

@article{tan2014magnetic,
  title={Magnetic hyperthermia properties of nanoparticles inside lysosomes using kinetic Monte Carlo simulations: Influence of key parameters and dipolar interactions, and evidence for strong spatial variation of heating power},
  author={Tan, Reasmey Phary and Carrey, Julian and Respaud, Marc},
  journal={Physical Review B},
  volume={90},
  number={21},
  pages={214421},
  year={2014},
  publisher={APS}
}

@article{ruta2015unified,
  title={Unified model of hyperthermia via hysteresis heating in systems of interacting magnetic nanoparticles},
  author={Ruta, Sergiu and Chantrell, R and Hovorka, O},
  journal={Scientific reports},
  volume={5},
  number={1},
  pages={9090},
  year={2015},
  publisher={Nature Publishing Group UK London}
}

@article{jonasson2019modelling,
  title={Modelling the effect of different core sizes and magnetic interactions inside magnetic nanoparticles on hyperthermia performance},
  author={Jonasson, Christian and Schaller, Vincent and Zeng, Lunjie and Olsson, Eva and Frandsen, Cathrine and Castro, Alejandra and Nilsson, Lars and Bogart, Lara K and Southern, Paul and Pankhurst, Quentin A and others},
  journal={Journal of Magnetism and Magnetic Materials},
  volume={477},
  pages={198--202},
  year={2019},
  publisher={Elsevier}
}

@article{suess2007reliability,
  title={Reliability of Sharrocks equation for exchange spring bilayers},
  author={Suess, D and Eder, S and Lee, J and Dittrich, R and Fidler, J and Harrell, JW and Schrefl, T and Hrkac, G and Schabes, M and Supper, N and others},
  journal={Physical Review B—Condensed Matter and Materials Physics},
  volume={75},
  number={17},
  pages={174430},
  year={2007},
  publisher={APS}
}

@incollection{gompper09,
  title = {Multi-Particle Collision Dynamics: {{A}} Particle-Based Mesoscale Simulation Approach to the Hydrodynamics of Complex Fluids},
  booktitle = {Advanced {{Computer Simulation Approaches}} for {{Soft Matter Sciences III}}},
  author = {Gompper, G. and Ihle, T. and Kroll, D. M. and Winkler, R. G.},
  editor = {Holm, Christian and Kremer, Kurt},
  year = 2009,
  series = {Advances in {{Polymer Science}}},
  number = {221},
  pages = {1--87},
  publisher = {Springer},
  address = {Berlin},
  doi = {10.1007/978-3-540-87706-6_1},
  isbn = {978-3-540-87705-9}
}

@article{mamiya2020estimation,
  title={Estimation of magnetic anisotropy of individual magnetite nanoparticles for magnetic hyperthermia},
  author={Mamiya, Hiroaki and Fukumoto, Hiroya and Cuya Huaman, Jhon L and Suzuki, Kazumasa and Miyamura, Hiroshi and Balachandran, Jeyadevan},
  journal={ACS nano},
  volume={14},
  number={7},
  pages={8421--8432},
  year={2020},
  publisher={ACS Publications},
  doi = {10.1021/acsnano.0c02521}
}

@article{kapral2008multiparticle,
  title={Multiparticle collision dynamics: Simulation of complex systems on mesoscales},
  author={Kapral, Raymond},
  journal={Advances in Chemical Physics},
  volume={140},
  pages={89},
  year={2008},
  publisher={Wiley Online Library}
}

@article{mostarac2025thermal,
  title={Thermal Stoner-Wohlfarth model for magnetodynamics of single domain nanoparticles: Implementation and validation},
  author={Mostarac, Deniz and Kuznetsov, Andrey A and Helbig, Santiago and Abert, Claas and S{\'a}nchez, Pedro A and Suess, Dieter and Kantorovich, Sofia S},
  journal={Physical Review B},
  volume={111},
  number={1},
  pages={014438},
  year={2025},
  publisher={APS}
}

@article{brown1963,
  title = {Thermal Fluctuations of a Single-Domain Particle},
  author = {Brown, William Fuller},
  journal = {Phys. Rev.},
  volume = {130},
  issue = {5},
  pages = {1677--1686},
  numpages = {0},
  year = {1963},
  month = {Jun},
  publisher = {American Physical Society},
  doi = {10.1103/PhysRev.130.1677},
  url = {https://link.aps.org/doi/10.1103/PhysRev.130.1677}
}

@article{poperechny2014dynamic,
  title={Dynamic hysteresis of a uniaxial superparamagnet: Semi-adiabatic approximation},
  author={Poperechny, IS and Raikher, Yu L and Stepanov, VI},
  journal={Physica B: Condensed Matter},
  volume={435},
  pages={58--61},
  year={2014},
  publisher={Elsevier},
  doi = {10.1016/j.physb.2013.08.049}
}

@article{weeks1971role,
	author = {Weeks, John D and Chandler, David and Andersen, Hans C},
	journal = {The Journal of chemical physics},
	number = {12},
	pages = {5237--5247},
	publisher = {AIP},
	title = {Role of repulsive forces in determining the equilibrium structure of simple liquids},
	volume = {54},
	year = {1971}}

@article{socoliuc_ferrofluids_2022,
        author = {Socoliuc, V. and Avdeev, M. V. and Kuncser, V. and Turcu, Rodica and Tomb{\'a}cz, Etelka and V{\'e}k{\'a}s, L.},
        date = {2022},
        doi = {10.1039/D1NR05841J},
        issn = {2040-3364, 2040-3372},
        journal = {Nanoscale},
        journaltitle = {Nanoscale},
        langid = {english},
        number = {13},
        pages = {4786--4886},
        shortjournal = {Nanoscale},
        shorttitle = {Ferrofluids and bio-ferrofluids},
        title = {Ferrofluids and bio-ferrofluids: looking back and stepping forward},
        url = {http://xlink.rsc.org/?DOI=D1NR05841J},
        urldate = {2022-04-19},
        volume = {14},
        year = {2022}}

@article{lundgren_dynamics_1983,
        title = {Dynamics of the {Relaxation}-{Time} {Spectrum} in a {CuMn} {Spin}-{Glass}},
        volume = {51},
        copyright = {http://link.aps.org/licenses/aps-default-license},
        issn = {0031-9007},
        url = {https://link.aps.org/doi/10.1103/PhysRevLett.51.911},
        doi = {10.1103/PhysRevLett.51.911},
        language = {en},
        number = {10},
        urldate = {2024-12-18},
        journal = {Physical Review Letters},
        author = {Lundgren, L. and Svedlindh, P. and Nordblad, P. and Beckman, O.},
        month = sep,
        year = {1983},
        pages = {911--914},
}

@article{hiroi_superspin_2011,
        title = {Superspin glass originating from dipolar interaction with controlled interparticle distance among $\gamma$-{Fe$_2$} {O$_3$} nanoparticles with silica shells},
        volume = {83},
        url = {https://link.aps.org/doi/10.1103/PhysRevB.83.224423},
        doi = {10.1103/PhysRevB.83.224423},
        number = {22},
        journal = {Physical Review B},
        publisher = {American Physical Society},
        author = {Hiroi, Kosuke and Komatsu, Katsuyoshi and Sato, Tetsuya},
        month = jun,
        year = {2011},
        pages = {36--9},
}

@article{materon2021magnetic,
  title={Magnetic nanoparticles in biomedical applications: A review},
  author={Mater{\'o}n, Elsa M and Miyazaki, Celina M and Carr, Olivia and Joshi, Nirav and Picciani, Paulo HS and Dalmaschio, Cleocir J and Davis, Frank and Shimizu, Flavio M},
  journal={Applied Surface Science Advances},
  volume={6},
  pages={100163},
  year={2021},
  publisher={Elsevier}
}

@article{wolfschwenger2025dual,
  title={Dual field magnetic separation for improved size fractionation of magnetic nanoparticles},
  author={Wolfschwenger, Manuel and Leliaert, Jonathan and Jaufenthaler, Aaron and Baumgarten, Daniel},
  journal={Nanoscale},
  pages ={23958--23970},
  year={2025},
  publisher={Royal Society of Chemistry}
}

@misc{InnMNP2026,
  author  = {Wolfschwenger, Manuel},
  title   = {InnMNP: Innsbruck Magnetic Nanoparticle Models},
  version = {1.0},
  year    = {2026},
  doi     = {10.5281/zenodo.20610900},
  url       = {https://doi.org/10.5281/zenodo.20610900},
  publisher = {Zenodo}
}

@article{pimk_FFBDMC,
        author = {Ilg, P and Kr{\"o}ger, M},
        journal = {Phys. Chem. Chem. Phys.},
        pages = {22244-22259},
        title = {Dynamics of interacting magnetic nanoparticles: effective behavior from competition between {B}rownian and {N}{\'e}el relaxation},
        volume = {22},
        year = {2020}}

@ARTICLE{abert_micromagnetics_2019,
       author = {{Abert}, Claas},
        title = "{Micromagnetics and spintronics: models and numerical methods}",
      journal = {European Physical Journal B},
         year = 2019,
        month = jun,
       volume = {92},
       number = {6},
          eid = {120},
        pages = {120},
          doi = {10.1140/epjb/e2019-90599-6},
       adsurl = {https://ui.adsabs.harvard.edu/abs/2019EPJB...92..120A}
}

@article{PhysRevE.106.064605,
  title = {Simulating the flow of interacting ferrofluids with multiparticle collision dynamics},
  author = {Ilg, Patrick},
  journal = {Phys. Rev. E},
  volume = {106},
  issue = {6},
  pages = {064605},
  numpages = {12},
  year = {2022},
  month = {Dec},
  publisher = {American Physical Society},
  doi = {10.1103/PhysRevE.106.064605},
  url = {https://link.aps.org/doi/10.1103/PhysRevE.106.064605}
}

@article{witt2005three,
  title={Three-dimensional micromagnetic calculations for naturally shaped magnetite: Octahedra and magnetosomes},
  author={Witt, Anne and Fabian, Karl and Bleil, Ulrich},
  journal={Earth and Planetary Science Letters},
  volume={233},
  number={3-4},
  pages={311--324},
  year={2005},
  publisher={Elsevier}
}

@misc{grad_2026_20450775,
  author       = {Grad, Jean-Noël and
                  Weik, Florian and
                  Reinauer, Alexander and
                  Kobayashi, Hideki and
                  Tischler, Ingo and
                  Blanco, Pablo M. and
                  Mostarac, Deniz and
                  Bindgen, Sebastian and
                  Beyer, David and
                  Hoßbach, Julian and
                  Kuron, Michael and
                  Hohenberger, Paul and
                  Müller, Niklas and
                  Frenner, Riccardo and
                  Gandhi, Yashas and
                  Brito, Mariano E. and
                  Tovey, Samuel and
                  Weeber, Rudolf},
  title        = {ESPResSo},
  month        = may,
  year         = 2026,
  publisher    = {Zenodo},
  version      = {5.0.1},
  howpublished = {Zenodo software release, version 5.0.1, DOI: 10.5281/zenodo.20450775},
  doi          = {10.5281/zenodo.20450775},
  url          = {https://doi.org/10.5281/zenodo.20450775},
  swhid        = {swh:1:dir:0c16b3a2f71a6a0646597d6aca1603f63f726f93
                   ;origin=https://doi.org/10.5281/zenodo.18791182;vi
                   sit=swh:1:snp:7b65dcfb7b815f912026e89a3d15801432b2
                   1cda;anchor=swh:1:rel:78e9cdc4b3b711523b02e738cf0a
                   e9f328003a85;path=espresso-5.0.1
                  },
}

\end{document}